\documentclass[%
 amsmath,amssymb,
 aps,prf,
]{revtex4-2}

\usepackage{graphicx}
\usepackage{dcolumn}
\usepackage{bm}
\usepackage{verbatim}
\usepackage[dvipsnames]{xcolor}

\begin{document}

\title{Complex segregation patterns in confined nonuniform granular shearing flows}


\author{Santiago Caro$^{1,2}$} 
 \email{Email: santiago.caro-alba@univ-eiffel.fr}
\author{Riccardo Artoni$^{1}$}
\author{Patrick Richard$^{1}$}
\author{Michele Larcher$^{2}$}
\author{James T. Jenkins$^{3}$}
\affiliation{$^1$MAST-GPEM, Université Gustave Eiffel, F-44344 Bouguenais, France}
\affiliation{$^2$Free University of Bozen-Bolzano, I-39100 Bozen-Bolzano, Italy}
\affiliation{$^3$Cornell University, Ithaca, New York 14053, USA}

\date{\today}


\begin{abstract}
\begin{color}{Black}
When polydisperse granular systems are sheared, the transverse dynamics is characterized by the interplay of size segregation and diffusion. Segregation in nonuniform and confined shearing flows is studied using annular shear cell experiments complemented with discrete numerical simulations of bidisperse, inelastic, and frictional spheres under gravity. We explored the role of shear localization, granular temperature, boundaries, and mixture properties in the evolution of the segregation rate and the maximum degree of segregation achieved by a bidisperse granular system in the steady state.
A faster segregation process and a more developed degree of segregation is observed for bidisperse mixtures with a larger size ratio and a higher proportion of large particles.
Normally, in the presence of gravity, size segregation induces large particles to rise and small particles to sink. However, two additional complex segregation patterns were found: inverse segregation and horizontal segregation. The first might be related to the kinematics of the flow, while the second is a geometrical effect. 
This additional segregation mechanism, in addition to diffusion fluxes and high confining pressure, hampers complete segregation in the steady state, where some degree of mixing always persists. 
\end{color}

 \end{abstract}
\maketitle
 \section{\label{sec:level1}Introduction}
 Flowing dense granular materials tend to segregate due to size polydispersity, which can be problematic in many industrial applications and natural hazards. In dense granular flows of particles of different sizes but same density, it has been observed that large particles tend to rise as small particles fall through voids under gravity, a segregation mechanism known as percolation or ``kinetic sieving'' \cite{Duan2021,Savage1988,Cooke1979, gray2018}.

 \begin{color}{Black}
 Size segregation in dense granular flows is known to be induced by pressure gradients \cite{Gray2005,gray2018,Fan2014,Jones2018,Duan2021,Singh2023} and shear rate gradients \cite{Duan2025, jing2021, Liu2023, Singh2023, Hill2008, Fan2010, Fan2011a, Xu2003, Louge2002}.
 In this regard, segregation can be characterized to involve two main driving mechanisms: gravity-related and flow kinematics-related \cite{Duan2025, jing2021, Liu2023, Singh2023}. 
 The pressure gradient, in the presence of gravity, and the shear rate gradient driving segregation can be counteracted by interspecies drag and diffusive remixing \cite{Duan2025, Gray2005}.
 \end{color}

 Size segregation has previously been studied using experimental and numerical approaches using different granular flow configurations \cite{gray2018}, for example: planar shear \cite{Scott1975,VanDerVaart2015,jing2021,Liu2023}, oscillatory shear \cite{,VanDerVaart2015}, annular shear \cite{Louge1996,Daniels2010,Tirapelle2021,Liu2023}, inclined plane or chute flow \cite{Savage1988,Gray2005,Wiederseiner2011,Tripathi2011,Thornton2012,Larcher2013,Larcher2015,DOrtona2020,Neveu2022,Liu2023,Singh2023}, heap flow \cite{Fan2014,Jones2018,Duan2020,Duan2021}, and spherical tumbler or rotating drum \cite{Ottino2000,DOrtona2016}.
 However, despite the number of studies, most of them dealt with uniform shearing and/or free surface flows. Few have addressed in detail the segregation behavior in nonuniform and confined shearing flows geometries \cite{Daniels2010, jing2021, Duan2025}, in which the rheology is known to become nonlocal, due to shear localization, creep and boundaries \cite{Artoni2018}.
 
 This work focuses on understanding segregation in nonuniform and confined shearing flows using annular shear cell experiments complemented with discrete numerical simulations of bidisperse, inelastic, and frictional spheres under gravity.
 Two complex segregation patterns are studied: inverse segregation, in which large particles concentrate in the bottom of the flow, where maximum shear and granular temperature are localized, instead of segregating to the top as expected; horizontal segregation, or accumulation of small particles next to the sidewalls. 
 Additionally, the role of the mixture properties is explored such as the size ratio and the mass fraction. 
 
 
 Previous studies of uniform shearing and/or free surface flows have shown that the tendency of spherical particles to sink or rise in a bidisperse mixture can be characterized by the size ratio between the diameter of the large and small particles $d_L/d_S$; and the mass fraction between the total mass of large and small particles in the mixture $M_L/M_S$ \cite{Gray2005, Duan2021, Fan2014, Tunuguntla2017, Jones2018, Duan2025}. Experimental works on vibratory die filling \cite{LAWRENCE1968,LAWRENCE1969} showed that segregation in flowing binary mixtures initially increases with a larger size ratio $d_L/d_S$, but plateaus in the interval $2.5<d_L/d_S<5$ and then decreases with further increases of $d_L/d_S$. 
 
 In terms of the relative species velocity, Discrete Element Method (DEM) simulations and experiments of free surface flows \cite{Jones2018, VanDerVaart2015, Duan2025} and confined shearing flows \cite{Duan2025}, describe asymmetric segregation velocities in bidisperse mixtures. 
 This means that a few small particles among many large particles segregate faster than few large particles among many small particles \cite{Jones2018}.
 Assuming mass conservation, the mass flux of small particles sinking downward must match that of large particles moving upward.
 The consequence is an asymmetry in the vertical flux at a low concentration of small particles, $C_S<0.5$ \cite{Jones2018,VanDerVaart2015, Jing2017}.
 
 Studies in quasistatic shear cells \cite{VanDerVaart2015} 
 or gravity-driven free surface flows \cite{Jones2018} have highlighted the direct correlation between the relative species velocity and the local volume concentration and the shear rate. This linear relative species velocity model \cite{Fan2014, Jones2018} depends on an empirical relation for the segregation length
 scale, and neglects the influence of pressure. Therefore, it is limited to free-surface flows where the effects of pressure gradients are negligible \cite{Duan2025}. 
 However, segregation in bidisperse mixtures is also affected by the amount of overburden pressure applied to the system \cite{Umbanhowar2019}. Experimental studies using both a split-bottom cell \cite{Hill2008} and an annular shear cell \cite{Golick2009} have demonstrated that applying confining pressure to the particle bed reduces both the speed at which segregation occurs and the overall degree of separation that develops. Similar trends have been reported in DEM simulations of horizontal planar shear flow, where increasing confining pressure was found to suppress segregation in several different cases: systems with a small fraction ($<0.1$ by volume) of fine particles mixed into a bed of larger ones \cite{Liu2017} and mixtures containing equal volumes of either size bidisperse particles \cite{Fry2018}.
 The latter have tried to make some adjustments to the linear relative species velocity model \cite{Fan2014, Jones2018}, so it can be applied in situations with significant overburden pressures \cite{Fry2018}. 
 \begin{color}{Black}
 More recent developments lead to relative species velocity models based on particle-level forces \cite{Duan2025} applicable to a full variety of flow and mixture conditions. The approach uses a force balance of the particle weight, the segregation force, and the granular drag force to determine the relative species velocity in mixtures of two particle species. It might include as well some corrections due to diffusion flux, as a consequence of concentration gradients, that can enhance or counteract the segregation flux \cite{Duan2025}. 
 
 Indeed, in infinite granular flows achieved in periodic streamwise geometries, the steady state has been described as the balance between segregation and diffusion fluxes \cite{Umbanhowar2019}. In this sense, in the steady state, defined at the moment in which the degree of segregation remains steady over time, there is always the coexistence of mixing and segregation \cite{Daniels2010}, which makes it not possible to have a granular system that segregates nearly completely.

 Additionally, in presence of gravity, size segregation generally induces large particles to the surface of the flow and small particles to concentrate at the bottom of the flow. However, in some cases it has been shown that large particles may partially segregate towards the bottom of the geometry (inverse segregation) \cite{Ortona2018}.
 \end{color}
 
 The annular shear geometry used in this work has been used to study monodisperse, nonuniform and confined flows \cite{Artoni2018}. In these experiments, particles are confined by a bumpy bottom rotating plate and a loaded bumpy top plate, while bounded by flat sidewalls. The streamwise velocity measured in this geometry follows an exponential depth profile. The shape of the profile will depend on the applied confinement pressure, independent of the shearing velocity. 
 From a qualitative point of view, the generated granular flows can be divided into two regions according to the streamwise velocity profiles: a lower shear zone, where most of the velocity variation is recorded (around $90\%$); and an upper creep zone, where the flow velocity is very slow \cite{Artoni2015a}. 
 Studies on density segregation performed in this nonuniform shearing geometry have highlighted that the segregation process is strictly related to the characteristics of the granular flow, and notably the velocity profile \cite{Tirapelle2021}. 

 The paper starts with a description of the experimental setup in Sec.~\ref{sec:Experimental set-up}, followed by the description of the numerical simulations carried out in Sec.~\ref{sec:Numerical simulations}.
 Subsequently, the experimental results are presented in Sec.~\ref{sec:Experimental results}, including the vertical segregation evolution at the sidewalls and the segregation profiles in the steady state, together with a description of the flow regime in experiments.
 The experimental results are then complemented with the corresponding numerical results in Sec.~\ref{sec:Numerical results}, including not only particles at the sidewalls but also particles inside the dense granular flows.




 

 \begin{figure}[h!]
  \centering
  \includegraphics{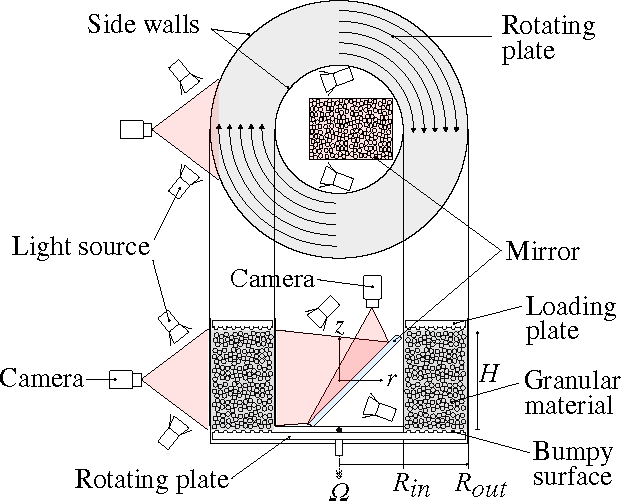}
  \caption{\label{fig:fig_Experiment_Sketch} Sketch of the experimental annular shear cell. Top view shows a bottom plate that rotates at an angular velocity $\Omega$. Cross section shows the granular material bounded by an outer ($R_\mathrm{out}$) and an inner ($R_\mathrm{in}$) sidewall, and confined by a top loading plate free to move up and down. $H$ is the depth of the flow.
  Granular flow is filmed from outer and inner sidewall using a mirror.}
 \end{figure}
%
 \section{\label{sec:Experimental set-up}Experimental setup}

  \subsection{\label{sec:level2}Experimental shear cell and conditions}

  Experiments on size segregation were performed in the bounded, nonuniform  annular shear configuration discussed above, and on which monodisperse flow properties and density segregation were characterized \cite{Artoni2018,Tirapelle2021} (see Fig.~\ref{fig:fig_Experiment_Sketch}). The vertical boundaries of the cell consist of two ﬂat, frictional, and transparent coaxial cylinders made of polymethyl methacrylate. 
  The inner and outer cylinders have diameters of 200 and 300 mm respectively, and both are 5 mm thick. This means that the flow annulus has an outer radius of $R_\mathrm{out}=145\, \mathrm{mm}$ and an inner radius of $R_\mathrm{in}=100\, \mathrm{mm}$, leaving an internal spacing of $\Delta R=R_\mathrm{out}-R_\mathrm{in}=45\,\mathrm{mm}$. The horizontal boundaries are two bumpy annular surfaces, which are made in polylactic acid 
  with an FDM three-dimensional printer to control the bumpiness. The bumpiness is achieved with hemispheres with diameter $d_W=4\,\mathrm{mm}$, distributed on a triangular lattice, spaced by a distance $s=2d_W$ between the centers. The bottom surface is glued to a rotating plate, driven by a brushless motor that allows to control the angular speed, which is kept fixed in this work at $\Omega = 18.5\,\mathrm{rpm}$. In order to compare with previous studies, it is useful to represent the velocity in dimensionless terms by means of the Froude number, considering the linear velocity in the middle of the annular spacing, as: 
  \begin{color}{Black}
  $\mathrm{Fr} = {\pi\Omega (R_\mathrm{in}+R_\mathrm{out})}/{60\sqrt{g d_S}} = 1.2$, 
  \end{color}
  where $g=9.81\,\mathrm{m/s^2}$ is the gravitational acceleration and $d_S=4\,\mathrm{mm}$ is the diameter of the small particle.
  The top wall is free to move in the vertical direction, allowing for volumetric changes as it applies a constant confining load. The top wall cannot rotate and therefore particles slip against this boundary.
 
 In this configuration, size-bidisperse mixtures of spheres were sheared. Polyoxymethylene (POM) ($\rho=1.41\,\mathrm{g/cm}^3$) spheres of diameter $4$, $6$, and $8\,\mathrm{mm}$ were used. 
 \begin{color}{Black}
 Polydispersity is practically negligible because for POM beads used it is equal or less than $\pm3\%$ of the diameter. 
 The diameter of the smallest particles $d_S=4\,\mathrm{ mm}$ was chosen as the length scale $d$ in the experiments.
 \end{color}
 The total amount of plastic beads in the cell was controlled by mass and was chosen to produce a granular flow characterized by a nonuniform shearing with a transition between a shear band and a creep region. Previous studies have shown that the thickness of the shear band is typically equal to $10d$ for monodisperse spheres \cite{Richard2020}. Therefore, it was decided to produce a granular flow deeper than $10d$, using a total particle mass of $M_L+M_S=2500\,\mathrm{ g}$, producing a granular flow approximately $H=25d$ deep. The confining top load was fixed to a value equal to the total particle mass $M_L+M_S$. 
 
 In order to test the effect of the mass fraction $M_L/M_S$ between large and small particles in the mixture, three different mixtures were tested: $M_L/M_S=$ 1/3, 1, and 3.
 Similarly, to address the effect of the particle size ratio, two combinations of particle diameters were tested: $d_L/d_S=$ 1.5 and 2. 
 
 The cell was initially filled following an inverse segregation pattern, with small particles layers over large particles layers.
 The mixtures were sheared for 30 min, long enough to reach the steady state of the segregation process in all experiments, where no further vertical migration of particles was observed at the sidewalls.
 
 \subsection{\label{sec:measurements_Exp}Flow properties measurements}
 
 Detailed sidewall measurements were acquired by filming the flow through the transparent sidewalls. Movies were recorded from outside, but also from inside, using a mirror installed inside the inner cylinder, as sketched in Fig.~\ref{fig:fig_Experiment_Sketch}. Two recordings were performed, one using a camera (Basler acA2440-75uc) at a frame rate of 1 fps for characterizing mixing and segregation during the whole experiment, and a second one 
 \begin{color}{Black}
 at the end of some experiments 
 \end{color}
 using a high-speed video camera (Phantom Miro 320S) at 200 fps to characterize the kinematics of the flow in the steady state. 
 
 The statistics of local granular configurations and motions were estimated from video sequences by postprocessing image analysis, using a particle tracking algorithm based on the ``Voronoï'' diagram \cite{Larcher2007, Capart2002}. 
 The algorithm determines the position and size of each particle visible from the outer and inner sidewalls of the annular shear cell during the entire experiment. 
 \begin{color}{Black}
 For each frame, the flow is divided into slices $\delta z \approx 2.5d$, and a spatial averaging of different quantities is performed for particles falling inside this averaging area. 
 
 To quantify the evolution of the segregation process, local mass concentrations were estimated for large particles $C_L$ and small particles $C_S$ in every slice as $C_i=\sum m_i$, where $m_i$ is the mass of a particle within the slice belonging to the species $i$. This value was normalized by the local total mass concentration in the slice, so $C_i/(C_L+C_S)$.
 \end{color}
 Collecting the evolution of all the instantaneous local mass concentrations of large particles $C_L/(C_L+C_S)$ and their spatial distribution, it is possible to track size segregation while the system is being sheared.

 For annular shear cell experiments, the evolution of the segregation process is measured in terms of shear deformation $\Delta x$, defined as the cumulative rotation distance of the bottom plate, measured in the middle of the annular spacing. At the end of the experiment the shear deformation was $\Delta x\approx400\times10^3 \mathrm{ mm}$ or $\Delta x\approx100\times10^3d$.
 
 In the steady state, depth profiles of the local mass concentration of large particles $C_L/(C_L+C_S)$ were obtained. Furthermore, at the end of selected experiments all particles were extracted by layers, then sieved, and 
 \begin{color}{Black}
 weighed
 \end{color}
 to determine the coarse local concentration of large particles $C_L/(C_L+C_S)$ inside the granular flow (see Fig.~\ref{fig:fig_Segregation_Profiles_Simu}).
 
 The flow properties were measured in the steady state and kinematic depth profiles were obtained for streamwise velocity $v_x$, solid fraction $\phi$, and granular temperature $T$. 
 First, by means of particle tracking algorithm mentioned above, particles streamwise velocity $v_x$ was measured. 
 Second, the solid fraction $\phi$ was derived from local grain patterns by comparing the area occupied by grains and the surrounding ``Voronoï'' cell.
 Third, the granular temperature $T$ was obtained from a detailed analysis of Lagrangian and Eulerian velocity correlations \cite{Larcher2007, Capart2002}. 
 The granular temperature is a measure of the velocity fluctuations of the particles \cite{Richard2020} and in this work it is defined as the weighted average of the temperatures of the two species of particles ($i,j$) \cite{Larcher2013},
  $T=(n_iT_i + n_jT_j)/(n_i + n_j)$. 
  For species $i$, $n_{i}$ is the number density, and
  $T_{i}= \frac{1}{3} \langle m_i V_i^2 \rangle$, where 
  $V_i$ is the magnitude of the particle velocity fluctuations, and $m_{i}$ is the particle mass. 

 \begin{color}{Black}
  The flow rheology was also characterized by the local inertial number $I$. At the grain level, this dimensionless quantity describes the kinematics of granular materials as a ratio between the local grains inertia and the global level of confinement: $I={\dot{\gamma}d}/{\sqrt{P/\rho_B}}$, where $\dot{\gamma}=\delta v_x/\delta z$ is the local shear rate obtained from the streamwise velocity $v_x$ profiles, $d$ is the mean or most repeated diameter of granular material particles, $P$ is the local confining stress estimated from the pressure gradient, and $\rho_B=\rho_p \langle \phi \rangle$ is the bulk density of the granular material obtained with the material density and the average solid fraction \cite{gdr2004}.
 \end{color}

 \begin{figure}
  \centering
  \includegraphics{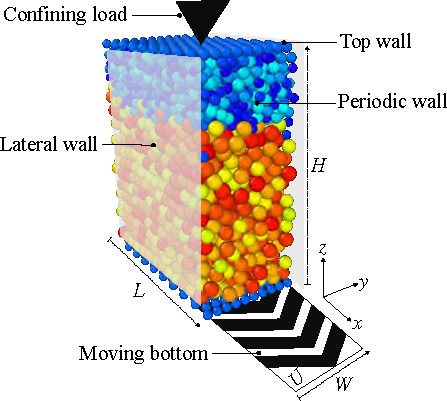}
  \caption{\label{fig:fig_Simulation_Lammps} Snapshot of the initial state of the simulated shear cell. Granular flow is bounded by two flat frictional lateral walls, separated at a distance of $W$, and confined by a top wall free to move up and down. $H$ is the depth of the flow. The bottom wall moves at a linear velocity $U$ in $x$ direction. 
  The cell is initially filled following an inverse segregation pattern (large particles at the bottom). 
  Colors correspond with particle size including 10\% polydispersity.}
 \end{figure}
 
 \section{\label{sec:Numerical simulations}Numerical simulations}

  \subsection{\label{sec:level2}Numerical method and setup}

 For investigating particle-scale phenomena, DEM simulations were performed \cite{Cundall1979,Luding2008}, using the open source LAMMPS platform \cite{LAMMPS2022}. The dense granular flows of the bidisperse mixtures were simulated using a periodic prismatic linear shear cell with a moving bumpy bottom, as previously done by Artoni and Richard \cite{Artoni2015a,Artoni2018,Artoni2021}. 
 
 All variables in the simulations are dimensionless. 
 \begin{color}{Black}
 The mean diameter of the small particles $d$ was chosen as the length scale. 
 \end{color}
 The mass $m$ of the particle with diameter $d$ was chosen as the mass scale. The timescale was chosen as $\sqrt{d/g}$, where $g$ is the gravitational acceleration. In practice, this corresponds to taking $d=1$, $m=1$ (or $\rho=6/\pi$), and $g=1$. 
 The purpose of the simulation is not to perfectly mirror experiments, but to understand the behavior of the particles inside the granular flow. Therefore, the simulation parameters are chosen with reasonable considerations, not calibrated.
 
 The geometry of the granular flow is a rectangular prism that scales the geometry of the experimental setup (see Fig.~\ref{fig:fig_Simulation_Lammps}), with a flow width of $W=11.25d$ and a length of $L=20d$. A periodic boundary condition is applied streamwise ($x$), which means that when a particle leaves the simulation box through a face, it is returned to the box, with the same velocity, at the opposite face. 
 
 The system studied is confined by two rough top and bottom walls, and two flat frictional sidewalls. The horizontal walls are made from 171 particles of diameter $d$, spaced in a triangular mesh at a distance slightly smaller than $2d$ between centers, ensuring that no particles can escape the simulation box. 
 
 \begin{color}{Black}
 The bottom wall moves at a constant dimensionless linear speed $U=2$, that can be expressed in terms of the Froude number $\mathrm{Fr}= U / \sqrt{gd}=2$. 
 \end{color}
 Shearing velocity in simulations is $67\%$ larger than the one used during the experiments. This increment responds to the need to reach the steady state sooner, due to computational time limitations, and since it has been shown that the depth velocity profile is independent of the shearing velocity \cite{Artoni2018}. 
 The top wall is free to move up and down and is responsible for confining the granular system with a force equal to the total weight of particles inside the cell, allowing the granular concentration to adjust. 
 
 The total mass of the particles was fixed to $M_L+M_S=6000\,m$, enough to fill the simulated volume and produce a granular flow that is approximately $H=25d$ deep. Similarly to experiments, the cell is initially filled with layers of small particles over layers of large particles (inverse segregation).
 
 Simulations were performed following the same systematic variations in the relative mass concentration of the species ($M_L/M_S=$ 1/3, 1, and 3), and the particle size ratio ($d_L/d_S=$ 1.5 and 2) used in the experiments, but introducing a slight polydispersity to avoid crystallization, ranging uniformly from $0.9d_i$ to $1.1d_i$ for each species $i$.

 The contacts between particles were simulated using a classical linear spring-dashpot model, with tangential elasticity and friction \cite{Silbert2001}, using the ``granular'' package in LAMMPS \cite{LAMMPS2022}.
 According to this model, the normal force between the particles $i$ and $j$ is given by $F_{n}^{ij}=k_{n}^{ij}\delta_{n}^{ij}-\gamma_{n} \dot\delta_{n}^{ij}$, where $k_{n}^{ij}$ is the normal stiffness, $\delta_{n}^{ij}$ is the particle interpenetration, and $\gamma_{n}=\gamma_{n_0} m_\mathrm{eff}^{ij} $ is the damping coefficient, where $m_\mathrm{eff}^{ij}=m_im_j/(m_i+m_j)$ is the effective mass of the interaction \cite{Shfer1996}. 
 To ensure the rigidity of the particles, the characteristic normal elasticity was set for the collision of two small particles $k_n^{SS}$, to produce a stiffness much greater than the estimated pressure $P$ at the bottom of the cell ${k_{n}^{SS}}/{P d_S}=10^{5}$. 
 The normal spring stiffness provides an intrinsic timescale for the system $t_{n}= \pi [{k_{n}^{ij}}/{m_\mathrm{eff}^{ij}}-(\gamma_{n}/2)^2] ^{-1/2} \simeq 1\times10^{-3} \sqrt{d/g}$ \cite{Shfer1996}. In DEM simulations, the computational time step is usually set as a small fraction of the collision time \cite{Artoni2021}. Thus, an integration time step of $1/100\mathrm{th}$ of a collision time was adopted $\Delta t=1\times10^{-5} \sqrt{d/g}$.
 For contacts involving at least one large particle, it was chosen to set the normal stiffness to yield the same collision time independently of the type of particles involved. This means, in practice, that all stiffnesses were set as follows: ${k_{n}^{SS}}/{m_\mathrm{eff}^{SS}}={k_{n}^{SL}}/{m_\mathrm{eff}^{SL}}={k_{n}^{LL}}/{m_\mathrm{eff}^{LL}}$. 
 Although it has been shown that the coefficient of restitution has nearly no influence on dense granular flows due to the presence of enduring contacts \cite{Dippel1999,Artoni2021,Artoni2015a},
 the viscous damping coefficient was adjusted to produce a normal restitution coefficient of $e_n=0.7$, increasing dissipation and ensuring inelastic particle-particle collisions \cite{Artoni2021}. The specific damping coefficient was therefore adjusted to this value of $e_n$.
 The tangential component of the force between the particles $i$ and $j$ is described using an elastic model with stiffness $k_t$ 
 and a frictional threshold $\mu$ \cite{Artoni2021}. 
 The tangential stiffness was set as $k_t/k_n= 2/7$, a classical assumption that ensures that the period of tangential oscillations is the same as the period of normal oscillations after a contact \cite{Artoni2021, Artoni2022}.
 The coefficient of sliding friction between a particle-particle contact was chosen to be $\mu_{pp}=0.5$. The relatively high value was chosen to increase dissipation and reach the steady state quicker \cite{Artoni2018}.
 The interaction of a particles with a flat wall was set to be inelastic and frictional, with a coefficient of friction $\mu_{pw}=0.3$. This value was chosen because previous studies have shown that larger values do not affect the velocity profile of the granular flow in this type of configuration \cite{Artoni2018}. 
 
 As will be discussed in the Sec.~\ref{sec:Numerical results}, the evolution of the segregation process is different in every simulation. Therefore, simulations were run until segregation reached a steady state for all the different mixtures. As explained in Sec.~\ref{sec:Experimental set-up}, the evolution of the flow is measured in terms of the total shear deformation $\Delta x/d$. The simulations were then run until $\Delta x\approx250\times10^3d$.  

 \subsection{\label{sec:Measurements_Simu}Flow properties measurements}

 Snapshots of particle positions and velocities are continuously saved from the simulations. Unlike experiments, all particle data are acquired, allowing for internal local measurements. Space averaging was done following the coarse-graining procedure explained by Artoni and Richard, using a Lucy weighting function with a compact support of $2d$ radius \cite{Artoni2015b,Artoni2018}.
 
 The segregation process was tracked by measuring the evolution of the local mass concentration of large particles $C_L/(C_L+C_S)$. 
 The evolution and degree of vertical segregation can also be represented in terms of the segregation index $\mathrm{SI}$, allowing a direct comparison between all the different mixtures. The segregation index is estimated as $\mathrm{SI}=2(\mathrm{COM}_L-\mathrm{COM}_S)/H$, where $\mathrm{COM}_L$ and $\mathrm{COM}_S$ are the vertical positions of the center of mass for large particles and small particles, respectively.
 This quantity varies from $-1$ (inverse segregation: all small particles above), to 1 (complete segregation: all large particles above), and 0 corresponds to homogeneous spatial distribution of particles \cite{DOrtona2020}.
 Size segregation in 
\begin{color}{Black}
 the steady state is quantified 
\end{color}
 with depth profiles of the local mass concentration of large particles, averaged over around 1000 integration time steps, after the shear deformation was $\Delta x \approx250\times10^3d$.

 Similarly, to characterize the flow in the steady state, depth profiles of the kinematic quantities (streamwise velocity $v_x$, solid fraction $\phi$, and granular temperature $T$) are obtained from the simulations in the same range of concentration profiles. 
\begin{color}{Black}
 The flow was also characterized by the local inertial number $I$ \cite{gdr2004}, following the same methodology described for the experiments in Sec.~\ref{sec:measurements_Exp}.
\end{color}

 The local vertical motion $v_z$ of particles species $i$ is known to depend from the local concentration \cite{Jones2018,VanDerVaart2015}. 
 Therefore, the relative species velocity $\omega_i= \langle v_{z}^{i} \rangle - \langle v_{z}^{B} \rangle$ was estimated for each species, where $\langle v_{z}^{i} \rangle$ is the local average velocity of a single species, and $\langle v_{z}^{B} \rangle = (\sum v_z^L\ C_L + \sum v_z^S\ C_S)/(C_L+C_S) $ is the local weighted average velocity of all species \cite{Jones2018}. 

 Additionally, to understand up to which point the flow could be collisional, collected contact forces were used to determine force chains networks and local average coordination number for all species $Z_L+Z_S=(Nc_L+Nc_S)/(Np_L+Np_S)$, where $Nc_i$ is the local number of contacts for species $i$, and $Np_i$ is the local number of particles of species $i$.
 

 \section{\label{sec:Experimental results}Experimental results}

  \begin{figure*}
  \centering
  \includegraphics{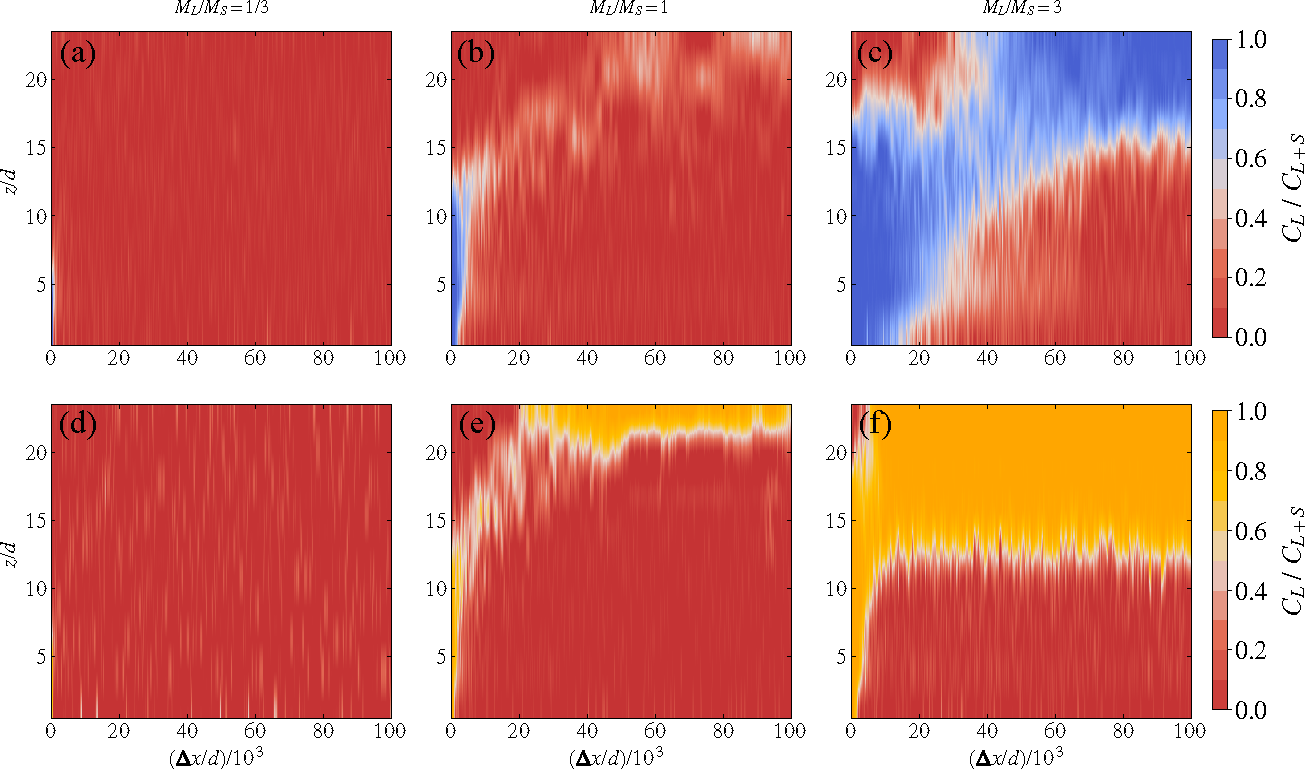}
  \caption{\label{fig:fig_Evolution_Vertical_Segregation_Experiments_Out} Evolution of the local concentration of large particles $C_L/(C_L+C_S)$ as a function of the dimensionless shear deformation $\Delta x/d$ in the outer sidewall of annular shear cell experiments. Small particles are represented in dark red, while large particles in midnight blue [(a)--(c)] and yellow [(d)--(f)]. Light colors represent a mixed state. Upper row [(a)--(c)] correspond to a particle size ratio of $d_L/d_S=1.5$, while bottom row [(d)--(f)] correspond to $d_L/d_S=2$. Columns from left to right correspond to mass fractions of $M_L/M_S=$ 1/3, 1, and 3, respectively.}
 \end{figure*}
 \begin{figure*}
  \centering
  \includegraphics{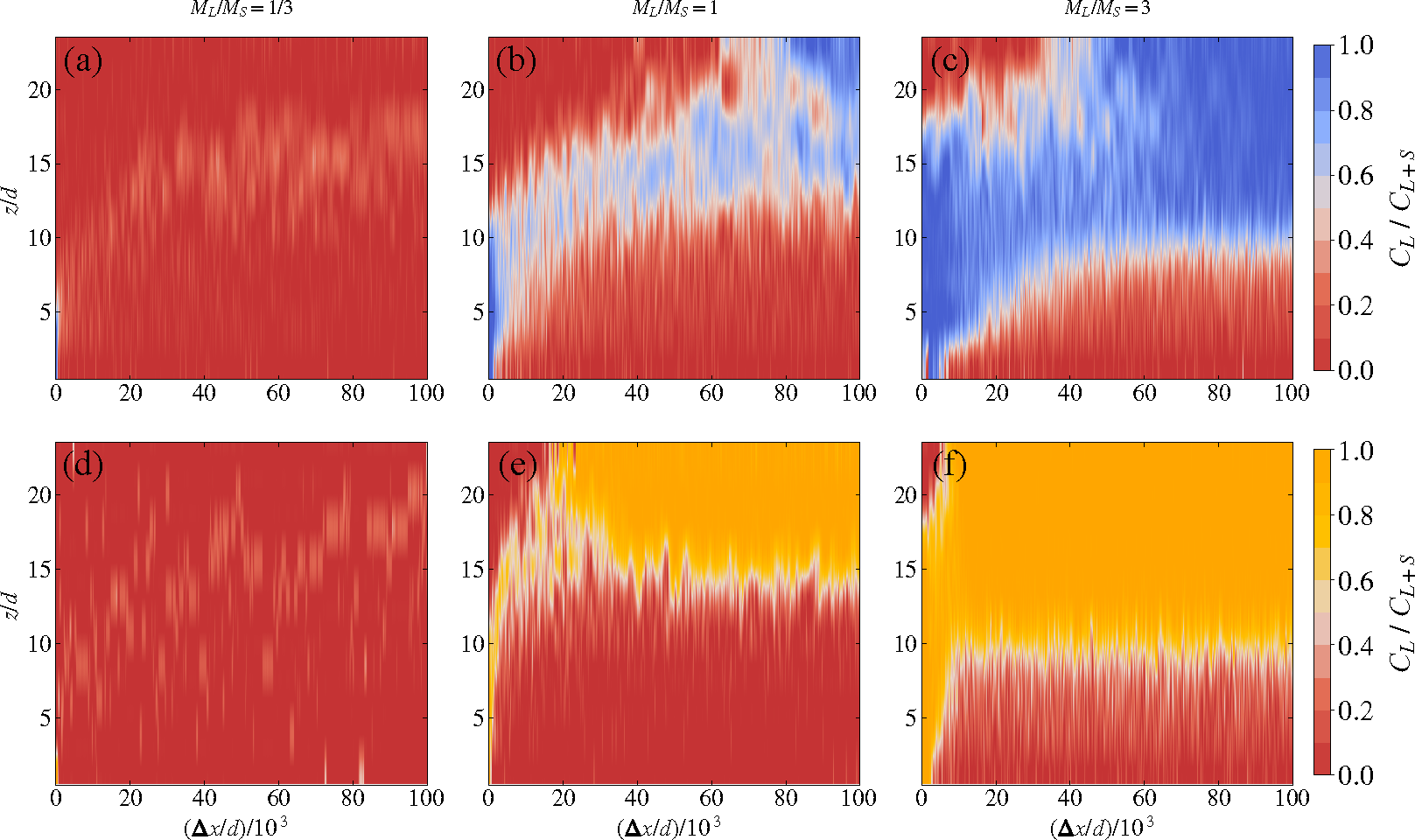}
  \caption{\label{fig:fig_Evolution_Vertical_Segregation_Experiments_In} Evolution of the local concentration of large particles $C_L/(C_L+C_S)$ as a function of the dimensionless shear deformation $\Delta x/d$ in the inner sidewall of annular shear cell experiments. Small particles are represented in dark red, while large particles in midnight blue [(a)--(c)] and yellow [(d)--(f)]. Light colors represent a mixed state. Upper row [(a)--(c)] correspond to a particle size ratio of $d_L/d_S=1.5$, while bottom row [(d)--(f)] correspond to $d_L/d_S=2$. Columns from left to right correspond to mass fractions of $M_L/M_S=$ 1/3, 1, and 3, respectively.}
 \end{figure*}

 \subsection{\label{sec:level2}Vertical segregation evolution at the sidewalls}
 
 The evolution of the local mass concentration of large particles $C_L/(C_L+C_S)$ is analyzed with color maps (Fig.~\ref{fig:fig_Evolution_Vertical_Segregation_Experiments_Out} and Fig.~\ref{fig:fig_Evolution_Vertical_Segregation_Experiments_In}). 
 \begin{color}{Black}
 From the maps, it is observed that in the systems initially inversely segregated, as shear is applied, segregation develops, making large particles rise and small particles sink, leading to a mixed state where particles are homogeneously distributed in space.
 After a certain shear deformation, specific for every system, segregation reaches a steady state, leaving some segregated regions, especially close to the top and bottom walls, and regions where some degree of mixing remains in between. 
 \end{color}
 
 Moreover, in proportion, more small particles are observed at the sidewalls than expected based on the mixture composition. It seems that small particles tend to accumulate at the sidewalls, pushing large particles inside the flow.
 These sidewalls observations suggest the presence of an additional horizontal segregation mechanism in this type of bounded granular flows. This is clear evidence that in this type of flow configuration, boundaries play an important role in the segregation process.
 For example, 
\begin{color}{Black}
packing fraction fluctuations as a function of the distance from the wall reflect the layering due to the order induced by the walls \cite{Artoni2015a, Camenen2013}. This order represents a drop in packing fraction due to larger interstitial voids close to the walls, compared to the inside of the flow \cite{gdr2004}. Smaller particles have higher probability of finding a suitable void in the layer close to the wall \cite{Schrter2006,Umbanhowar2019}. The probability increases with larger size ratios, as interstitial voids are bigger between larger particles. Once a layer of small particles is formed at the wall, the voids left between small particles and wall are too small to fit a larger particle, excluding larger particles to the core of the flow and constituting an alternative segregation mechanism.

\end{color}

 Something particular is observed in experiments. The segregation pattern at the outer sidewall (see Fig.~\ref{fig:fig_Evolution_Vertical_Segregation_Experiments_Out}) differs from that observed at the inner sidewall (see Fig.~\ref{fig:fig_Evolution_Vertical_Segregation_Experiments_In}). In the inner sidewall, vertical segregation is more developed compared to the outer sidewall. 
 The reason for this difference is attributed to a geometrical effect and is further discussed in Sec.~\ref{sec:segregation_Profiles_Exp} when analyzing the concentration profiles obtained from inner and outer walls.  

 When comparing the different concentration maps obtained from the inner and outer sidewalls, it is confirmed that the evolution of vertical segregation depends on the mass fraction of the species $M_L/M_S$ and the particle size ratio $d_L/d_S$ \cite{Gray2005, Duan2021, Fan2014, Tunuguntla2017, Jones2018}.
 For higher values of $M_L/M_S$ and $d_L/d_S$, the segregation process is faster and more developed, consistent with previous experimental studies \cite{LAWRENCE1968,LAWRENCE1969}. 
 Additionally, the smaller $M_L/M_S$, the larger the number of small particles that tend to accumulate near the sidewalls, keeping away large particles from the sidewalls, making it difficult to track size segregation. 

  \begin{figure}
  \centering
  \includegraphics{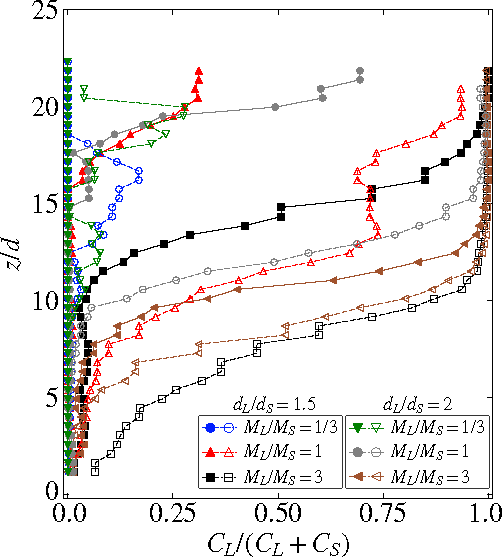}
  \caption{\label{fig:fig_Segregation_Profiles_Experiments} Depth profiles of the mass concentration of large particles $C_L/(C_L+C_S)$ in the steady state ($\Delta x\approx100\times10^3d$) in annular shear cell experiments. Continuous lines with filled markers correspond to data from the outer sidewall, while dashed lines with void markers correspond to data from the inner sidewalls.}
 \end{figure}
 \begin{figure*}
   \centering
   \includegraphics[width=\textwidth]{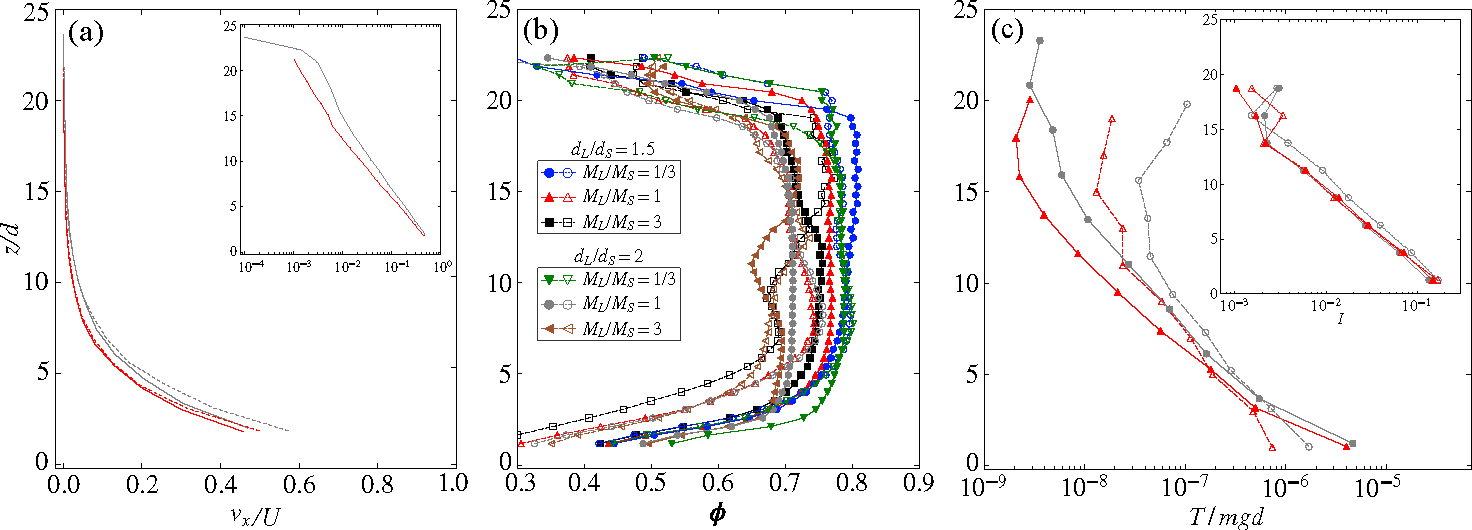}
   \caption{\label{fig:fig_Kinematics_Experiments} Average kinematic depth profiles measured at the shear cell sidewalls in the steady state ($\Delta x\approx100\times10^3d$) in experiments. Continuous lines correspond to outer sidewall, while dashed lines to inner sidewalls. (a) Streamwise velocity $v_x$ rescaled by the velocity $U$ of the bottom wall (inset represents the $x$ axes in logarithmic scale). (b) Solid fraction $\phi$. (c) Mixture granular temperature $T/\mathrm{(mgd)}$ (inset represents the corresponding inertial number $I$).
   }
 \end{figure*}

  \subsection{\label{sec:segregation_Profiles_Exp}Vertical segregation in the steady state at the sidewalls }
  
 To assess the overall balance between mixing and segregation, once the steady state was reached, depth profiles of the local concentration of large particles $C_L/(C_L+C_S)$ were produced using data from the inner and outer sidewalls (see Fig.~\ref{fig:fig_Segregation_Profiles_Experiments}). These concentration profiles evidence the difference between the segregation patterns observed in the inner and outer sidewalls. 

 \begin{color}{Black}
 The mechanism behind the different segregation patterns observed at the sidewalls may be related to a geometric effect due to the curvature and concavity of the sidewalls or also the commensurability of the circumference and the particle diameter, leading to different packing levels, as discussed in Sec.~\ref{sec:kine_exp}, and therefore different relative concentrations.
 \end{color}
 
 In terms of the different mixture properties, higher degrees of segregation are reached for larger values of the mass fraction $M_L/M_S$ and the particle size ratio $d_L/d_S$, on both the inner and outer sidewalls. 
 When comparing the concentration profiles obtained at the inner and outer sidewalls, segregation patterns can be grouped according to the mass fraction $M_L/M_S$. 
 Starting with the mixtures where $M_L/M_S=1/3$, it can be seen that for the outer sidewall the value of $C_L/(C_L+C_S)\approx0$ along the whole depth, while at the inner side the concentration of large particles increases to $C_L/(C_L+C_S)=0.25$ above $z=10d$. 
 For mixtures where $M_L/M_S=1$, the inner and outer sides correspond qualitatively but the segregation is more developed at the inner sidewall.
 Now, for mixtures where $M_L/M_S=3$, similar smooth sigmoidal concentration profiles are observed at the outer and inner sidewalls. The only difference is the degree of segregation or the number of large particles that successfully segregate to the top of the flow. 
 It can be concluded that segregation can be underestimated in the outer sidewall due to the accumulation of small particles at the sidewall; however, this does not mean that the whole segregation process is represented from the inner sidewall because there is also the effect of the horizontal segregation mechanism.


 \subsection{\label{sec:kine_exp}Flow characterization: Kinematic profiles in the steady state}

 At the end of the experiments, kinematic quantities were measured at the inner and outer sidewalls to characterize the granular flow, as described in Sec.~\ref{sec:Experimental set-up}. 
 Kinematic quantities were then averaged locally along the depth of the granular flow $z/d$ to obtain depth profiles (see Fig.~\ref{fig:fig_Kinematics_Experiments}).
 Additionally, the range of the inertial number $I(z/d)$ in which the granular systems are being deformed is included.
  \begin{color}{Black}
 To optimize the overall time of the experimental campaign, it was decided to obtain representative velocity measurements only from the equal mass fraction condition.
 \end{color}
 
 The average streamwise velocity $v_x$ profiles [see Fig.~\ref{fig:fig_Kinematics_Experiments}(a)] show that particles in experiments are sheared with a nonuniform shear rate, without evident differences between the inner and outer sidewalls. The $v_x$ profiles follow the shape of an exponential function \cite{Artoni2018}, where $v_x$ is maximum at the bottom of the flow and decreases nonlinearly toward the top of the flow, where $v_x$ is close to zero. 
 The maximum $v_x$ registered corresponds to about $70\%\mbox{--}80\%$ of the imposed shear velocity of the rotating bottom plate. 
 Now, based on the $v_x$ profiles, an arbitrary distinction is made to split the flow into two main regions \cite{Artoni2015a}: a shear zone, from the bottom to approximately $z=10d\mbox{--}12d$, in which most of the shear takes place and where the velocity decays by 90\% and a creep zone above, in which the particles move as a plug and where the velocity is minimum.

 Measurements of the solid fraction $\phi$ occupied by particles at the outer and inner sidewalls (see Fig.~\ref{fig:fig_Kinematics_Experiments}b) show a loose region near the moving bottom ($z=0d$) where the system experiences maximum shear. The mixtures become better packed as the shear decreases toward the top of the flow, far from the moving bottom. Close to the top wall, $\phi$ drops again to values similar to those at the bottom of the flow. The value of $\phi$ becomes almost constant above a depth of $z \approx 7d$ when $d_L/d_S=1.5$ and above a depth of $z \approx 4d$, when $d_L/d_S=2$. 
 \begin{color}{Black}
 In general, the high measured solid fraction values are atypical for assemblages of spheres but normal for assemblages of disks with some ordering. This is consistent for 2D measurements obtained from the sidewalls.
 \end{color}

 In the steady state, the packing measured at the sidewalls of the annular shear cell depends on $M_L/M_S$ and $d_L/d_S$ and may differ between the sidewalls.
 \begin{color}{Black}
 Mixtures where $M_L/M_S=1/3$, achieved the best packed configurations, due to the accumulation of only small particles that order and crystallize at the sidewalls due to the low polydispersity. 
 \end{color}
 Furthermore, $\phi$ measured at the outer and inner sidewalls of these mixtures does not show marked differences in the steady state. 
 
 In contrast, mixtures in which $d_L/d_S=1.5$ and $M_L/M_S\geq1$, show different values of $\phi$ between outer and inner sidewalls. 
 Particles at the outer concave sidewall $R_\mathrm{out}$ pack better than those at the inner convex sidewall $R_\mathrm{in}$, at least below $z \approx 12d$ where only small particles are found in the steady state. 
 \begin{color}{Black}
 Better packing results in a higher concentration of small particles and fewer chances to see a large particle at the outer concave sidewall compared to the inner convex sidewall.
 \end{color}

 The measured granular temperature profiles of the mixture $T$ [see Fig.~\ref{fig:fig_Kinematics_Experiments}(c)] show maximum values near the moving bottom of the flow and decrease exponentially to the top. Near the moving bottom wall, a value of $T \approx 10^{-1}\,\mathrm{mgd}$ is measured. Then $T$ decreases to a minimum of $T \approx 10^{-4}\,\mathrm{mgd}$ at $z \approx 15d$ for all mixtures.
 Below $z \approx 5d$, the temperature profiles $T$ outside are higher than those measured inside. Above this depth, the trend is inversed and the $T$ profiles inside become higher.
 The temperature profiles $T$ inside and outside show similar behavior in mixtures with different size ratios $d_L/d_S$.
 The profiles become constant above $z \approx 10d$ for the inner profiles, while for the outer profiles it takes up to $z \approx 17d$.
 Note that the behavior of the temperature, according to the distance from the bed, is the opposite of that of the concentration fraction in Fig.~\ref{fig:fig_Segregation_Profiles_Experiments}.

 \begin{color}{Black}
 To better understand the local flow conditions, especially when compared to other dense granular flows, the inset in Fig.~\ref{fig:fig_Kinematics_Experiments}c represents the corresponding inertial number $I$ estimated for some experiments. The range of inertial numbers, $I=[10^{-3}\mbox{--}10^{-1}]$, corresponds to a dense flow in which particle contacts are important \cite{gdr2004}.
 \end{color}
 
 Overall, experimental confined shearing flows tested in this work can be characterized by shear and temperature profiles nearly independent of $d_L/d_S$. This may be due to the fact that, due to size segregation, the bottom part is composed primarily of small particles, which determine the properties of the shear zone. 
 In this work, the effect of the mass fraction $M_L/M_S$ on the kinematics of the flow was explored with numerical simulation in Sec.~\ref{sec:kine_simu}.
 
 \begin{color}{Black}
 The solid fraction is affected by the degree of segregation achieved by a mixture, which depends as well on the mixture properties.
 The measured packing in experiments can be very high because of crystallization at the sidewalls due to low polydispersity in POM particles of the same species.
 
 In terms of the inner and outer sidewalls observations, no significant differences were observed in the shear and temperature profiles. However, a strong difference was seen between the inner and outer solid fractions.
 For some mixtures, it is clear that particles at the inner wall are more dilute and disordered. This is probably the reason why fewer small particles are seen at the inner sidewall. For a geometrical reason, particles do not pack very efficiently at the inner wall, which allows larger voids, resulting in more space for large particles.

 \end{color}

 \section{\label{sec:Numerical results}Numerical results}

 \begin{figure*}
   \centering
   \includegraphics{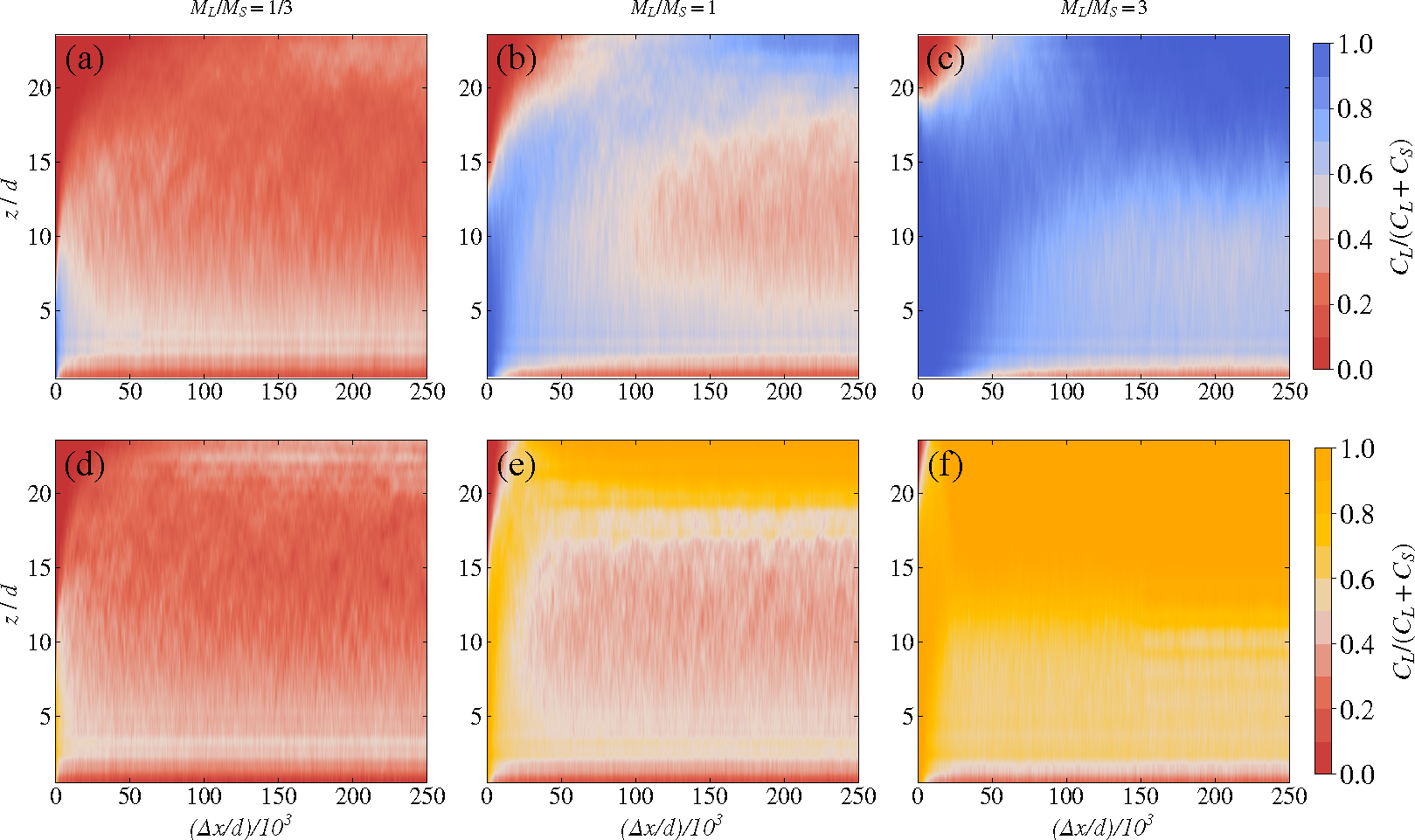}
   \caption{\label{fig:fig_Evolution_Vertical_Segregation_Simu} Evolution of the local concentration of large particles $C_L/(C_L+C_S)$ as a function of the shear deformation $\Delta x/d$ in numerical simulations. Small particles are represented in dark red, while large particles in midnight blue [(a)--(c)] and yellow [(d)--(f)]. Light colors represent a mixed state. Upper row [(a)--(c)] correspond to a particle size ratio of $d_L/d_S=1.5$, while bottom row [(d)--(f)] correspond to $d_L/d_S=2$. Columns from left to right correspond to mass fractions of $M_L/M_S=$ 1/3, 1, and 3, respectively.}
 \end{figure*}
\begin{figure}
\centering  
\includegraphics{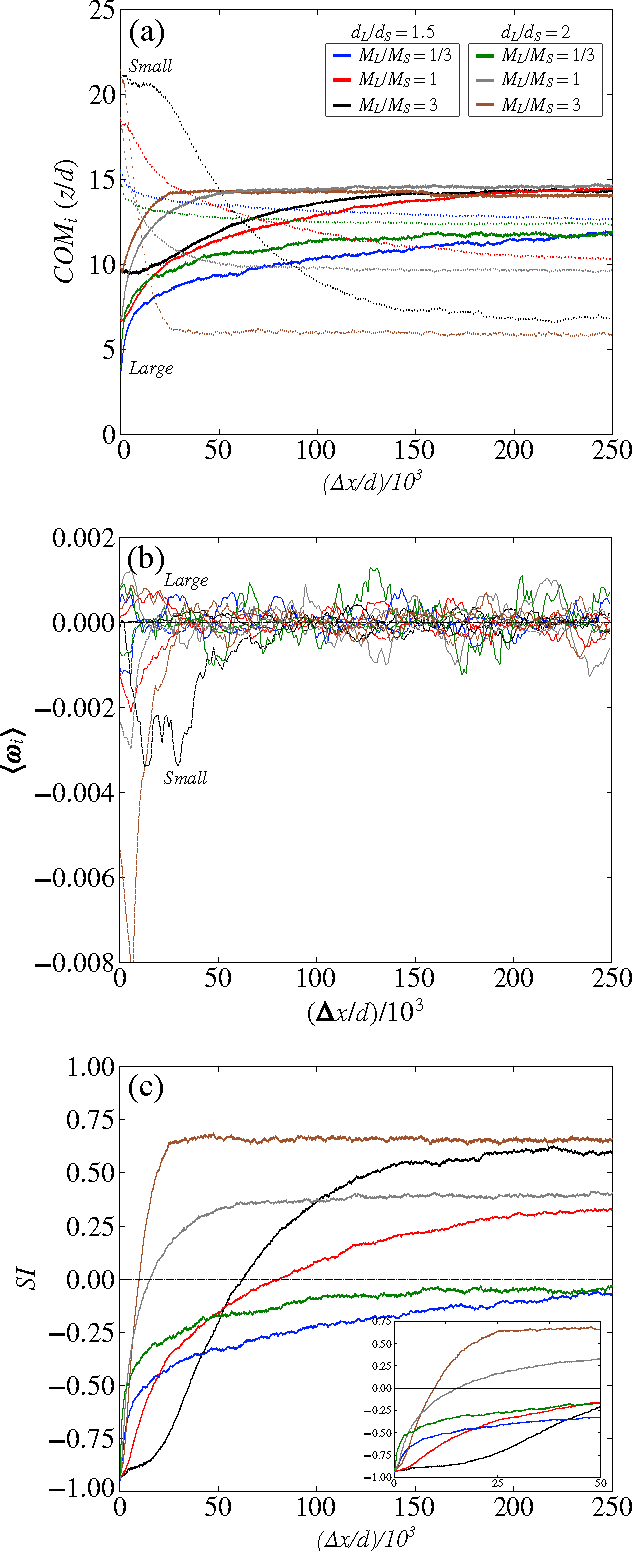}
\caption{\label{fig:fig_cmi_wi_SegregationIndex} (a) Evolution of the vertical center of mass of large (continuous line) and small (dotted line) particles $\mathrm{COM}_i\ (z/d)$.
(b) Evolution of the mean relative species velocity $\langle \omega_i \rangle$ of large (continuous line) and small (dashed line) particles.
(c) Evolution of the vertical segregation index $\mathrm{SI}$. Inset represents the very early behavior.
All data correspond to numerical simulations.}
\end{figure}
 
 \subsection{\label{sec:level2}Vertical segregation evolution}
  
 During simulations, the position of each particle was saved in time, mapping the evolution of the segregation process in terms of the local mass concentration of large particles $C_L/(C_L+C_S)$ at different depths $z/d$ of the granular flow (see Fig.~\ref{fig:fig_Evolution_Vertical_Segregation_Simu}). 

\begin{color}{Black}
 Similar to experimental observations, it is observed that in the systems initially inversely segregated, as shear is applied, segregation develops, making large particles rise and small particles sink, leading to a mixed state where particles are homogeneously distributed in space.
 After a certain amount of shear deformation, specific for every mixture configuration, segregation reaches a steady state, leaving some segregated regions, especially close to the top and bottom walls, and regions where some degree of mixing remains in between.
 In fact, some larger particles may concentrate at the bottom of the granular flow, even after considerable shear deformation, evidence of an additional inverse segregation mechanism.    
 \end{color}
 
 When comparing the different configurations, it is clear that the segregation rate and the degree of segregation reached by a mixture depend on the mass fraction $M_L/M_S$ \cite{Gray2005, Duan2021, Fan2014, Tunuguntla2017, Jones2018} and the particle size ratio $d_L/d_S$ \cite{Gray2005, Duan2021, Fan2014, Tunuguntla2017, Jones2018}. 
 This means that by increasing the fraction of larger particles in the mixture and/or the size ratio between the particles, segregation reaches a higher degree for a smaller shear deformation $\Delta x/d$ \cite{LAWRENCE1968,LAWRENCE1969}. 

 Another way to track segregation is to look at the overall vertical position of particles according to their size, also known as the center of mass $\mathrm{COM}_i$ where $i$ is the particle species (see Fig.~\ref{fig:fig_cmi_wi_SegregationIndex}a).
 \begin{color}{Black}
 When comparing the evolution of the center of mass $\mathrm{COM}_i$ of large and small particles, it is clear that most of the large particles rise, and most of the small particles sink.
 At the beginning of the shearing, the vertical total mass fluxes are maximum. The initial maximum mass fluxes respond to the initial segregated condition, where local concentrations are extreme. Then the mass fluxes decrease as the particles segregate and local concentrations become less extreme and more homogeneous. When the mass of every species is evenly distributed along the depth of the flow, the centers of mass of both species intercept at the center of the flow $z=H/2\approx12.5d$, indicating a homogeneous spatial distribution of particles. 
 From there, mass fluxes keep decreasing until the center of mass $\mathrm{COM}_i$ reach a plateau simultaneously for small and large particles, indicating that during size segregation the total mass flux of large particles is balanced by the total mass flux of small particles.

 When comparing different mixtures and species, if the mass fraction $M_L/M_S=1$, the displacement of the center of mass $COM_i$ is symmetric between large and small particles; however, this is not the case when the mass fraction $M_L/M_S\neq1$, because the total mass of the species are different, the species velocity must adjust to ensure that the total mass fluxes are balanced. When $M_L/M_S=3$ small particles need to globally segregate 3 times faster compared to large particles and vice versa when $M_L/M_S=1/3$. 
 
 The global efficiency of the segregation process increases with a larger size ratio $d_L/d_S$ and a higher mass fraction $M_L/M_S$, which means reaching a more developed degree of segregation (final displacement of the $\mathrm{COM}_i$) in the steady state, for a smaller shear deformation $\Delta x/d$.
 The final position of the center of mass of small particles $\mathrm{COM}_S$ is controlled mainly by $M_L/M_S$ and $d_L/d_S$. In contrast, the final positions of the centers of mass of large particles $COM_L$ seem to be limited for some unknown reason at $z\approx15d$. 
 Additionally, reaching lower degrees of segregation when the mass fraction is $M_L/M_S=1/3$ compared to $M_L/M_S=3$ indicates an asymmetry in the total mass fluxes. In both mixtures, the relative mass fraction of particles is equivalent, but the total mass fluxes do not correspond. In other words, mixtures with few small particles among many large particles $M_L/M_S=3$ are more efficient segregating compared to mixtures with few large particles among many small particles $M_L/M_S=1/3$.
 \end{color}

 To verify the hypothesis of different segregation velocities for large and small particles depending on the size ratio $d_L/d_S$ and mass fraction $M_L/M_S$, the evolution of the mean relative species velocity $\langle \omega_i \rangle$ was tracked in the shear zone for each species [see Fig.~\ref{fig:fig_cmi_wi_SegregationIndex}(b)]. As explained in Sec.~\ref{sec:Measurements_Simu}, the relative species velocity $\omega_i$ represents the difference between the local average relative velocity of each species, and the local weighted average relative velocity of all species. 
 \begin{color}{Black}
 These local values were averaged along the depth of the flow and some moving averaging was applied to reduce the strong fluctuations of the instantaneous species velocities.
 
 The evolution of the average relative species velocity is characterized by an initial spike on the onset of the flow, negative for small particles, and positive for large particles. However, the positive and negative spikes do not compensate the difference in mixture composition to ensure mass fluxes balance. Small particles spikes are higher than large particles spikes on the onset of the flow, even for mixtures where equal presence of both species $M_L/M_S=1$ and mixtures dominated by small particles $M_L/M_S=1/3$. This indicates that small particles may segregate without corresponding segregation of large particle. 
 The spikes take place before the centers of mass of both species intercept, consistent with initial maximum mass fluxes due to the initial extreme local relative mass concentrations because particles are initially segregated. 
 This initial state is rather extreme, with a pure phase of the smaller sphere above a pure phase of the larger. Segregation initially occurs only in the neighborhood of the interface between the two types of sphere. Here, with the spheres interacting through collisions, smaller spheres are more mobile and can first penetrate between and under the larger spheres to then force the movement of the larger upward. This is reflected in the initial downward excursion of the smaller spheres relative velocity.
 This phenomenon typically increases with size ratio $d_L/d_S$ due to voids size, and the mass fraction $M_L/M_S$ that controls the amount of large particles beneath small particles.

 After the initial spike, the average relative species velocity tends to zero and increments on the relative velocity of large particles seem to be corresponded by changes in the relative velocity of small particles. 
 Then, segregation velocities keep fluctuating around 0, meaning that species keep changing their sign. In mixtures where segregation is well developed $M_L/M_S>1/3$, this inversion on the sense of segregation is mainly observed after the centers of mass $\mathrm{COM}_i$ had reached a steady state.
 This might reflect the expected balance between segregation and diffusion fluxes, where some degree of mixing always persists \cite{Umbanhowar2019}.

 Previous works have addressed the dependency of the relative species velocity $\omega_i$ on the local relative concentration of small particles $C_S/(C_L+C_S)$ \cite{Jones2018,VanDerVaart2015,Duan2025,Jing2017}. 
 As shown in Fig. S1 of the Supplemental Material in the dense flows tested here this is not so clear and data is too noisy~\cite{Caro_SuppMat_2026}. The relative species velocity was scaled by the shear rate $\omega_i/\dot \gamma\ \delta z$ (see Fig. S3 in Supplemental Material~\cite{Caro_SuppMat_2026}) since this approach has been widely used \cite{Jones2018};
 Alternatively, the square root of the granular temperature was used $\omega_i/\sqrt{T}$ (see Fig. S4 in Supplemental Material~\cite{Caro_SuppMat_2026}) but did not scale well neither.
 A possible explanation for this could be the confined nature of the granular flows tested in this work, but also the evolution in space and time of our systems, passing throughout a transient regime until reaching a fully developed steady state.
 Strong fluctuations of the instantaneous species velocities are therefore natural, ending up with positive and negative velocities for the same species. 

 A better scaling was observed by taking the absolute value of the relative species velocity $|\omega_i|$ (see Fig. S5 and Fig. S6 in Supplemental Material ~\cite{Caro_SuppMat_2026}) that is a sort of measure of the intensity of the fluctuations, and therefore a better scaling was obtained with the granular temperature.  
 Additionally, the correlation between the shear rate and the square root of the granular temperature (see Fig. S2 in Supplemental Material) is not linear and there is not collapse. Therefore, it is possible that the $z/d$ dependence of the square root of the temperature is missed by the shear rate scaling.

 \end{color}

 In this work, the efficiency of the segregation process is indicated by the maximum degree of segregation reached by a mixture and the amount of shear deformation needed. Therefore, to quantify the efficiency of segregation and assess the effect of the mass fraction $M_L/M_S$ and the particle size ratio $d_L/d_S$ on size segregation, a new parameter was introduced: the segregation index $\mathrm{SI}$. As explained in Sec.~\ref{sec:Measurements_Simu}, $\mathrm{SI}$ represents the distance between the center of mass of small and large particles \cite{DOrtona2020}. 
 The evolution of $\mathrm{SI}$ as the systems are sheared [see Fig.~\ref{fig:fig_cmi_wi_SegregationIndex}(c)] shows that all mixtures start from a inversely segregated state ($\mathrm{SI}\approx-1$). In this initial configuration, the particles are well segregated and the distance between the center of mass is maximum ($\approx 12.5d$ or $\approx H/2$). 
 Then $\mathrm{SI}$ grows at different rates depending on $M_L/M_S$ and $d_L/d_S$, until it reaches a maximum degree of segregation that also depends on $M_L/M_S$ and $d_L/d_S$. 
 In general, segregation is more efficient in mixtures with higher values of $M_L/M_S$ and $d_L/d_S$, which means a higher degree of segregation for a smaller shear deformation. 
 \begin{color}{Black}
 However, looking in more detail how $\mathrm{SI}$ grows at the beginning of every simulation [see Fig.~\ref{fig:fig_cmi_wi_SegregationIndex}(c), inset], it was found that for systems with lower $M_L/M_S$ and higher $d_L/d_S$ segregation is initially faster; also, there is a delay in the beginning of the segregation process, and the effect is more notorious for larger $M_L/M_S$ and smaller $d_L/d_S$; this can be explained due to the extreme initial configuration of the particles, in addition to the mass fraction and size ratio that control the relative species velocity at the onset of the flow, as discussed above.
 \end{color}

\subsection{\label{sec:Vertical segregation in the steady state}Vertical segregation in the steady state} 

  \begin{figure}
   \centering
   \includegraphics{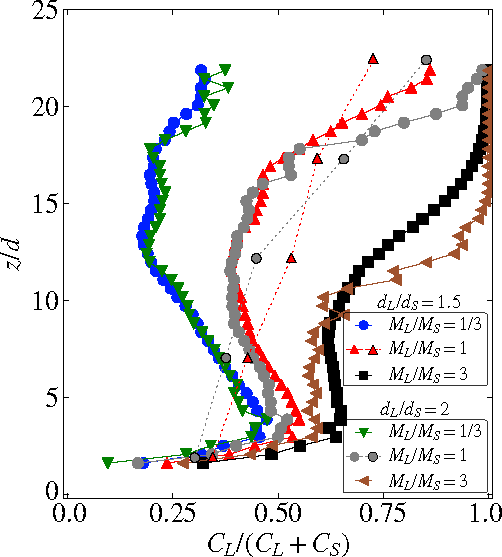}
   \caption{\label{fig:fig_Segregation_Profiles_Simu} Depth profiles of the mass concentration of large particles $C_L/(C_L+C_S)$ in the steady state ($\Delta x/d\approx250\times10^3d$) in numerical simulations. Dashed lines represent two representative experiments where all particles where extracted and counted at the end of the experiment ($\Delta x\approx100\times10^3d$).}
 \end{figure}

 Local mass concentration profiles of large particles $C_L/(C_L+C_S)$ were obtained for all simulations in the steady state, counting all particles inside the flow (see Fig.~\ref{fig:fig_Segregation_Profiles_Simu}).
 The profiles are characterized by a ``\textsf{S}'' shape. The local concentration of large particles $C_L/(C_L+C_S)$ is minimum at the bottom of the flow for all mixtures, between 0.10 and 0.30. Moving upward in the flow, $C_L/(C_L+C_S)$ grows to a local maximum at $z\approx4d$ for all mixtures, reaching values between 0.45 and 0.65. Above this depth, the profiles diverge according to $M_L/M_S$ and nearly independently of $d_L/d_S$. 
 For mixtures where most of the particles are small $M_L/M_S=1/3$, $C_L/(C_L+C_S)$ decreases up to $z\approx13d$. Then $C_L/(C_L+C_S)$ grows again, reaching a second concentration peak $C_L/(C_L+C_S)\approx0.3$ at $z\approx23d$, smaller than the maximum concentration recorded close to the bottom of the flow.
 For mixtures in which $M_L/M_S=1$, $C_L/(C_L+C_S)$ decreases to $\approx$0.4 at $z\approx13d$. Then $C_L/(C_L+C_S)$ grows again, reaching a global maximum of $C_L/(C_L+C_S)\approx0.9$ at a depth $z\approx23d$.
 For mixtures where most of the particles are large $M_L/M_S=3$, $C_L/(C_L+C_S)$ remains almost constant $\approx 0.65$ up to $z\approx10d$. Then $C_L/(C_L+C_S)$ grows again, reaching maximum values of the local concentration of large particles $C_L/(C_L+C_S)=1$ above $z\approx15d$.
 These observations confirm that the degree of segregation reached by a mixture depends on the mass fraction $M_L/M_S$ and the size ratio $d_L/d_S$. For higher values, segregation is more efficient and a higher fraction of particles sort by size.

  \begin{figure}
   \centering
   \includegraphics{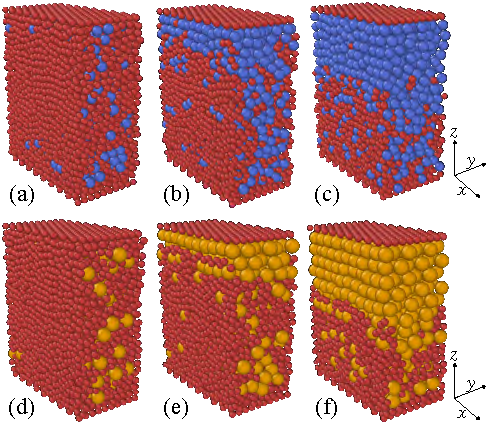}
   \caption{\label{fig:fig_t11000Snapshots_Simulations} Snapshots from numerical simulation in the steady state ($\Delta x\approx250\times10^3d$). Small particles are represented in dark red, while large particles in midnight blue [(a)--(c)] and yellow [(d)--(f)]. Upper row [(a)--(c)] correspond to a particle size ratio of $d_L/d_S=1.5$, while bottom row (d-f) correspond to $d_L/d_S=2$. Columns from left to right correspond to mass fractions of $M_L/M_S=$ 1/3, 1, and 3, respectively.} 
  \end{figure}
 \begin{figure*}
   \centering
   \includegraphics[width=\textwidth]{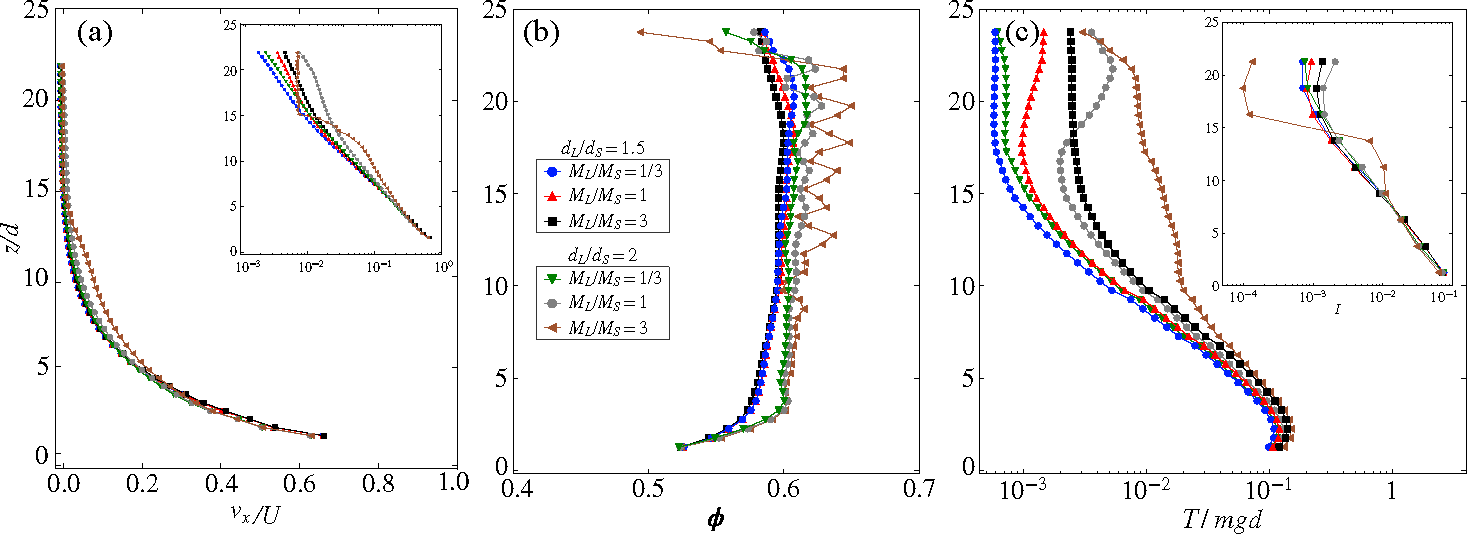}
   \caption{\label{fig:fig_Kinematics_Simulations} Average kinematic depth profiles obtained in the steady state ($\Delta x\approx250\times10^3d$) in numerical simulations. (a) Streamwise velocity $v_x$ rescaled by the velocity $U$ of the bottom wall (inset represents the $x$ axes in logarithmic scale). (b) Solid fraction $\phi$. (c) Granular temperature $T/\mathrm{mgd}$ (inset represents the corresponding Inertial number $I$).
   }
 \end{figure*}
 
 \begin{color}{Black}
 An interesting inverse segregation feature is conserved in all concentration profiles. In the presence of gravity, large particles accumulate not only at the top of the flow but also at the bottom of the flow, specifically below the half of the flow ($z<12d$). 
 This has been observed in previous simulations and experiments in which large particles might find equilibrium at an intermediate depth depending mainly on the size ratio \cite{Ortona2018}. For the size ratios used here ($d_L/d_S<2$), large particles are expected to go to the top of the flow, as most do; however, few of them concentrate toward the bed of the flow.
 Nevertheless, there is some evidence of large particles segregating in regions of higher shear rate and high granular temperature, in shearing flows in the absence of gravity \cite{Fan2011, Duan2025}.
 \end{color}
 The concentration of large particles $C_L/(C_L+C_S)$ at the bottom of the flow increases with the mass fraction $M_L/M_S$, and can be even higher than the concentration registered in the creep zone, as when $M_L/M_S=1/3$.
 
 Furthermore, as explained in Sec.~\ref{sec:measurements_Exp}, to obtain comparable concentration profiles for the experiments, at the end of selected experiments, all particles were extracted by layers to determine the local concentration of large particles $C_L/(C_L+C_S)$ as a function of the flow depth 'see Fig.~\ref{fig:fig_Segregation_Profiles_Simu} dashed lines). This contributes to overcome the limited data available from the sidewalls in experiments, which are not enough to describe the size segregation process inside the flow, due to boundary effects. 
 The experimental profiles obtained for the local concentration of large particles $C_L/(C_L+C_S)$ align well with the general trend shown by the concentration profiles in the simulations.
 However, the experimental profiles are less detailed and do not represent well the inverse segregation feature observed in the simulations.
 Therefore, in experiments, another segregation mechanism tending to accumulate large particles in the shear zone should be analyzed by a more accurate probing of the inner distribution of particles, which should be performed by noninvasive techniques, such as x-ray tomography \cite{Ando_2021}, for example.

 To visualize the maximum degree of segregation reached by different mixtures in the steady-sate, 3D snapshots of the simulated shear cell are included (see Fig.~\ref{fig:fig_t11000Snapshots_Simulations}). The snapshots show that for a higher value of the mass fraction $M_L/M_S$, more large particles reach the top layer of the granular flow (see Fig.~\ref{fig:fig_t11000Snapshots_Simulations} left-right). Snapshots also help to visualize that for different values of the size ratio $d_L/d_S$, the maximum degrees of segregation reached are similar (see Fig.~\ref{fig:fig_t11000Snapshots_Simulations} top-bottom). These snapshots also show a perspective of every configuration, where there is a clear accumulation of small particles close to the walls (horizontal segregation), as seen in experiments; but also the inverse segregation feature, accumulation of large particles close to the bottom of the flow.


\subsection{\label{sec:level2}Flow characterization in the steady state}
\subsubsection{\label{sec:kine_simu}Kinematic profiles}

\begin{figure*}
   \centering
   \includegraphics{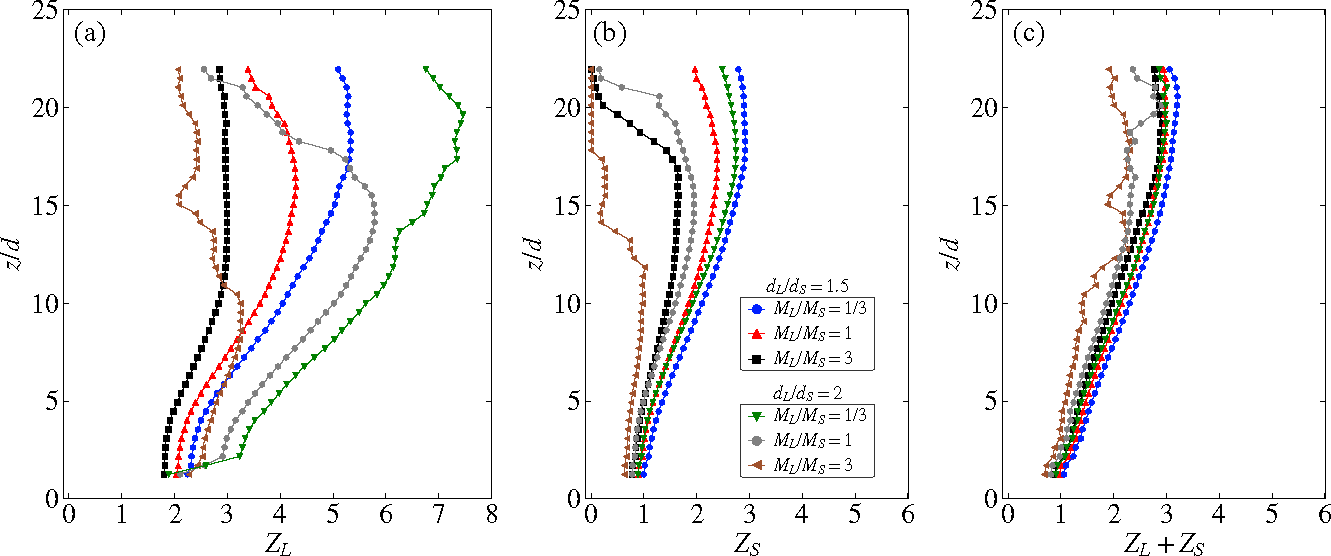}
   \caption{\label{fig:fig_CN_profile} Local coordination number $Z$ depth profile in the steady state ($\Delta x\approx250\times10^3d$) obtained in numerical simulations. (a) Local coordination number for large particles $Z_L$. (b) Local coordination number for small particles $Z_S$. (c) Local average coordination number for all species $Z_L+Z_S$.}
\end{figure*}

 Once the steady state was identified for all simulations, kinematic quantities were obtained as described in Sec.~\ref{sec:Measurements_Simu}. 
 The kinematic quantities were then averaged locally along the depth of the granular flow $z/d$ to obtain depth profiles (see Fig.~\ref{fig:fig_Kinematics_Simulations}).
 Additionally, the range of inertial number $I(z/d)$ at which the granular systems are being deformed is included. 
 %
 The first kinematic property studied is the streamwise velocity $v_x$. Streamwise depth velocity profiles are built from locally averaging the instantaneous streamwise velocity data from all particles in the steady state [see Fig.~\ref{fig:fig_Kinematics_Simulations}(a)]. 
 The velocity profiles appear to be independent of the mass fraction $M_L/M_S$ and the particle size ratio $d_L/d_S$ in the steady state.
 The exponential shape of the velocity profiles \cite{Artoni2018} shows that in simulations particles are sheared with a non-uniform shear rate, replicating the shearing configuration observed in the experiments. The profiles show that the velocity $v_x$ is maximum at the bottom of the flow and decreases nonuniformly toward the top of the flow, where it is close to zero. 
 Numerical data show that particles at the bottom slip and flow at about $70\%-80\%$ of the imposed shear velocity $U$.
 As in experiments, based on the $v_x$ profiles, an arbitrary distinction is made to divide the flow into two regions \cite{Artoni2015a}: a shear zone, from the bottom to approximately $z\approx10d-12d$, where a transition zone appears, leading to a creep zone above $z>12d$. 
 
 The average solid fraction $\phi$ occupied by the particles was obtained as a function of the flow depth $z/d$ [see Fig.~\ref{fig:fig_Kinematics_Simulations}b]. Starting from the bed, a loose region is characterized near the moving bottom where the system experiences maximum shear. The mixtures become more well packed as the shear decreases toward the top of the flow, far from the moving bottom. 
 The solid fraction $\phi$ becomes almost constant above a depth of $z \approx 10d$ when the particle size ratio is $d_L/d_S=1.5$; and above $z\approx4d$ for a particle size ratio $d_L/d_S=2$. Mixtures with a higher size ratio $d_L/d_S$ achieve a higher packing value. Some fluctuations between the values of $\phi=0.60\mbox{--}0.65$ are observed above $z\approx 10d$ for simulations where $M_L/M_S=3/1$. These systems are characterized by the layering of large particles on top of the granular flow [see Figs.~\ref{fig:fig_t11000Snapshots_Simulations}(c) and 
\ref{fig:fig_t11000Snapshots_Simulations}(f)]; 
  the fluctuations originate from this ordered structure, partially smoothed out by the coarse graining width used.

 The depth profiles of the granular temperature of the mixture $T$ were obtained as explained in Sec.~\ref{sec:Measurements_Simu}. The resulting profiles [see Fig.~\ref{fig:fig_Kinematics_Simulations}(c)] share nearly the same shape with some differences in magnitude depending on the mass fraction $M_L/M_S$ and the particle size ratio $d_L/d_S$. The mixture temperature profiles show a minimum near the bottom of the flow. Then temperature increases to a maximum at $z \approx 3d$ for all mixtures. This suggest that the bottom wall is dissipative, and that the higher velocity fluctuations are located in the shear zone. Above this depth, $T$ decreases for all mixtures. Around $ \approx 10d$ there is a change in the trend, and all temperature profiles diverge. 
 Around $z \approx 18d$, the mixture temperature $T$ becomes constant when $M_L/M_S=1/3$, while it tends to increase again when $M_L/M_S=1$. When $M_L/M_S=3$, $T$ becomes constant for $d_L/d_S=1.5$; however, when $d_L/d_S=2.0$, $T$ continuously decreases from $z \approx 10d$ to $z \approx 25d$, showing the greater velocity fluctuations.

 \begin{color}{Black}
 The corresponding inertial numbers $I(z/d)$ were estimated for all simulations [see Fig.~\ref{fig:fig_Kinematics_Simulations}(c), inset]. The range corresponds to the one found for the experiments, $I=[10^{-3}\mbox{--}10^{-1}]$, characterizing the simulated flows as dense flows in which the particles are in continuous contact \cite{gdr2004}.
 \end{color}

 Globally, simulations of confined shearing flows tested in this work show that kinematic quantities are nearly independent of the mixtures properties in the steady-sate, specially the shear rate and solid fraction.
 Differences in the granular temperature profiles may be explained due to different segregation patterns.
 In the shear zone the local mass concentration is similar for the different mixtures (see Fig.~\ref{fig:fig_Segregation_Profiles_Simu}), so similar granular temperatures are found. In the creep zone compositions depends more on mixture properties, so do granular temperature. The higher the concentration of large particles on the creep zone, the higher is the granular temperature in the creep zone. 
 
 \subsubsection{\label{sec:level2}Particles interactions}
 
 The local mean coordination number $Z_L+Z_S$ quantifies the average number of contacts or interactions made by a particle at the same time. 
 This quantity was then averaged locally along the depth of the granular flow $z/d$ to obtain depth profiles in the steady state [see Fig.~\ref{fig:fig_CN_profile}(c)].
 The results show that the local mean coordination number $Z_L+Z_S$ is minimum at the bottom of the flow, increases toward the top of the flow, and decreases again near the top of the flow for all simulations.  
 
 Typically, to form a force chain, a single particle must be in contact with at least two other particles at the same time ($Z=2$). In simulations, the local mean coordination number is $Z_L+Z_S<2$ below a depth of $z\approx10d$ [see Fig.~\ref{fig:fig_CN_profile}(c)]. This corresponds to the arbitrary division of the flow between the shear zone and the creep zone made based on the streamwise velocity profiles [see Fig.~\ref{fig:fig_Kinematics_Simulations}(a)]. Therefore, the shear zone can also be characterized as a nearly collisional region with a network of less connected force chains.

 Analyzing the local coordination number of large $Z_{L}$ and small $Z_{S}$ particles separately, it was found that $Z_{L}$ [see Fig.~\ref{fig:fig_CN_profile}(a)] is overall higher than $Z_{S}$ [see Fig.~\ref{fig:fig_CN_profile}(b)]. This can be explained by the bigger surface of larger particles.
 The coordination number of small particles $Z_{S}$ depends on the mass fraction $M_L/M_S$ and the particle size ratio $d_L/d_S$. In contrast, the coordination number of large particles $Z_{L}$ does not show a clear dependence on these two variables.

 \begin{color}{Black}
 After characterizing the simulated flow, it seems that the accumulation of large particles with depth in the shear zone is linked with looser and more agitated regions because the solid fraction and the coordination number decrease with depth in the shear zone, as well as with increasing shear rate and granular temperature with depth, including a point where the granular temperature profile changes its concavity due to the presence of the bottom boundary. 
 \end{color}
\section{\label{sec:level1}Conclusions}

Bounded and confined granular shearing flows of size-bidisperse mixtures were studied to determine the coupling of the size segregation process with the flow and mixture properties. 
Two techniques were used: annular shear cell experiments and DEM linear shear cell simulations.
Both approaches are designed to apply constant shearing at the bottom of a granular mixture, bounded with flat and frictional sidewalls, and confined with a load on top.
Both experiments and simulations are not identical, but complement each other to build a complete characterization of the flow properties and the segregation mechanism, since sidewall observation in experiments fail to represent all the segregation mechanisms inside the flow.
In total, six different granular mixtures were tested for two different size ratios and three different mass fractions.
Larger values of the particle size ratio and the mass fraction of large particles increase the efficiency of size segregation, reflected in higher segregation rates and degrees of segregation. In other words, bidisperse systems with fewer small particles are more efficient segregating, and even more for a bigger size difference between small and large particles.

Kinematics observed in experiments and simulations correspond qualitatively in the steady state, and are nearly independent of the mixture properties, with the exception of the granular temperature that seems to correspond to segregation patterns. 
Streamwise velocity profiles describe a nonuniform shearing flow with a shear band about $10d$ thick, located at the bottom of the flow, followed by a creep zone above.
The shear zone is characterized by larger shear-rate gradients, higher granular temperature, lower solid fraction, and fewer force chains, compared to the upper creep zone where enduring contacts maintain an active network of force chains.


\begin{color}{Black}
It was observed that in the steady state some degree of mixing always persists so that segregation is never complete, as evidence on the concentration profiles for the absence of a sharp transition. The main reason is the balance between segregation and diffusion fluxes, in addition to the high confining pressure.
Moreover, in the steady state, two additional segregation mechanisms are observed: large particles concentrating and layering close to the bottom of the flow, where maximum shear rates are present; and small particles accumulating near the sidewalls. The first one, corresponds to an inverse segregation mechanism in which some large particles find an intermediate equilibrium depth in the shear zone;
the second one, can respond to a geometric effect that promotes ordering and lower packing fractions near sidewalls, combined with the higher probability of small particles to occupy the interstitial voids leave between large particles and walls. 
\end{color}



To go further in the understanding of these complex flows, it is necessary to study in more detail the two complex segregation mechanisms presented. For this purpose, it is necessary to characterize the flow properties according to the distance from the sidewalls and the vertical position within the granular flow. It is necessary to study the presence of secondary recirculating flows, as well as confirm the effect of the confining pressure on size segregation patterns in nonuniform granular flows.
It could be interesting to explore the effect of the initial arrangement of particles on the size segregation rates and patterns observed in the steady state.
Additionally, granular flows in the absence of gravity may help characterize size segregation mechanisms under only shear rate gradients. 
It could also be interesting to study size segregation in multidisperse mixtures in confined shearing flows.                                               
In terms of methods, there is also a wide spectrum of advances in experimental techniques such as x-rays or refractive-index-matched materials that might help to see and understand segregation inside the granular flows. 






\begin{acknowledgments}
This project has received funding from the European Union's Horizon 2020 research and innovation programme under the Marie Skłodowska-Curie COFUND Grant Agreement No 101034248.
\end{acknowledgments}

\clearpage

\bibliography{Caro2025}

\end{document}


\title{Supplemental Material for: Complex segregation patterns in confined nonuniform granular shearing flows}
\maketitle


\begin{figure}[H]
   \centering
   \includegraphics{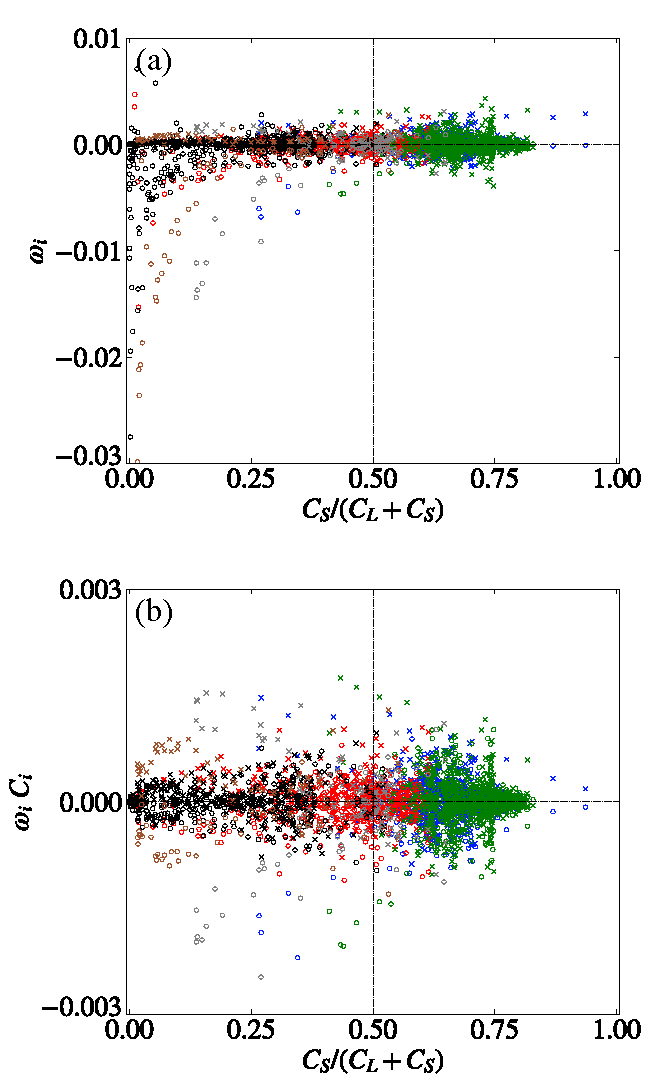}
   \caption{\label{fig:Suplementary_S1} Relative species velocity $\omega_i$ (a) and flux $\omega_i C_i$ (b) as a function of the local small particles concentration $C_S/(C_L+C_S)$ in numerical simulations.
   Data correspond only to the shear zone $z<12.5d$ before the steady state. Colors represent different simulations, large particles are represented with ``x'' markers, and small particles with ``o'' markers. }
 \end{figure}

  \begin{figure}[H]
   \centering
   \includegraphics{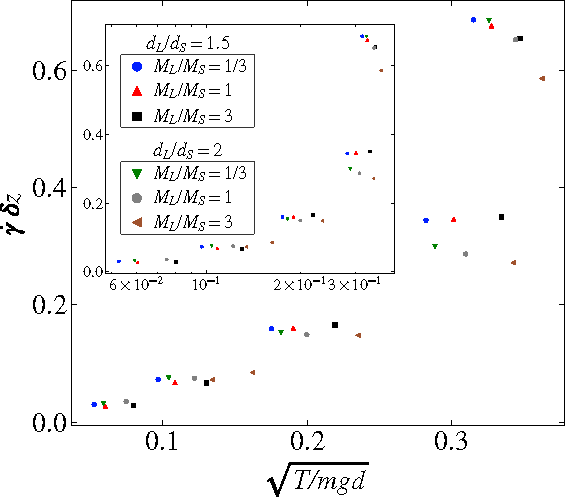 }
   \caption{\label{fig:Temperature_vs_Shear} Correlation between the shear rate $\dot \gamma\ \delta z$  and the square root of the granular temperature $\sqrt{T/mgd}$ in numerical simulations (inset represents the $x$-axes in logarithmic scale). Data correspond only to the shear zone $z<12.5d$ in the steady state.}
 \end{figure}

 \begin{figure}[H]
   \centering
   \includegraphics{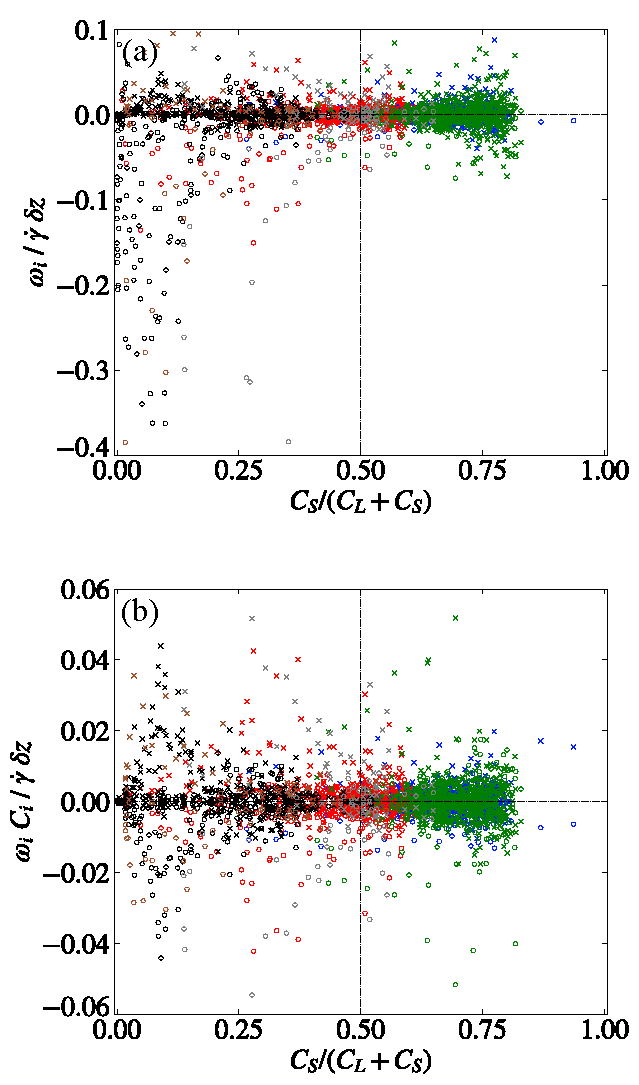 }
   \caption{\label{fig:Suplementary_S2} Relative species velocity $\omega_i$ (a) and flux $\omega_i C_i$ (b), scaled by the local shear rate $\dot \gamma\ \delta z$, as a function of the local small particles concentration $C_S/(C_L+C_S)$ in numerical simulations.
   Data correspond only to the shear zone $z<12.5d$ before the steady state. Colors represent different simulations, large particles are represented with ``x'' markers, and small particles with ``o'' markers. }
 \end{figure}

\begin{figure}[H]
   \centering
   \includegraphics{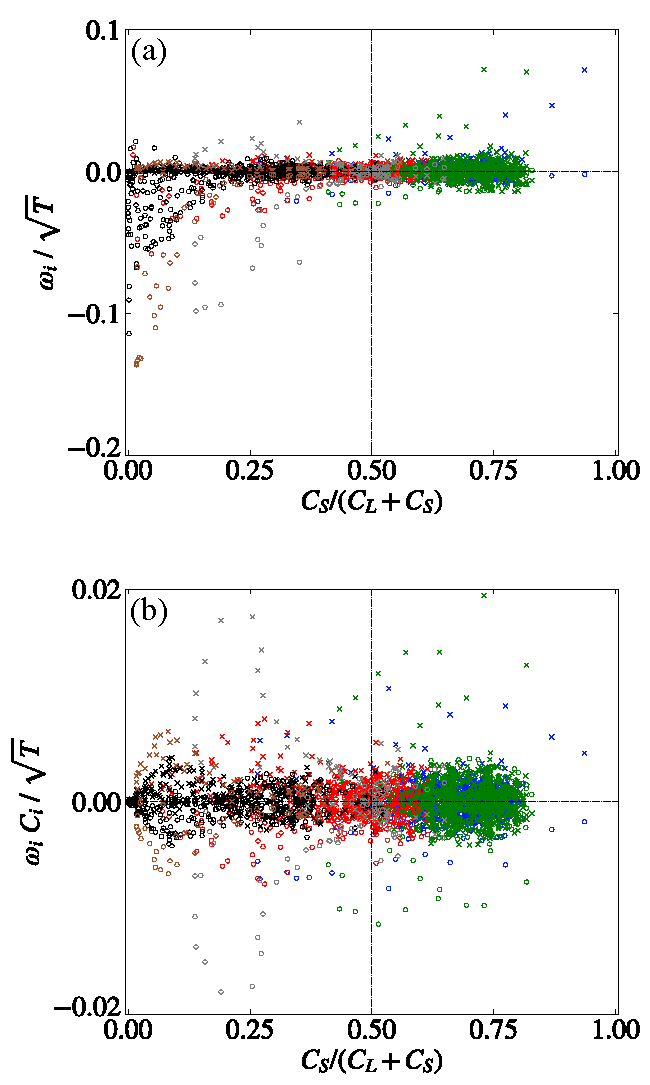 }
   \caption{\label{fig:Suplementary_S4} Relative species velocity $\omega_i$ (a) and flux $\omega_i C_i$ (b), scaled by the square root of the granular temperature $\sqrt{T}$, as a function of the local small particles concentration $C_S/(C_L+C_S)$ in numerical simulations. 
   Data correspond only to the shear zone $z<12.5d$ before the steady state. Colors represent different simulations, large particles are represented with ``x'' markers, and small particles with ``o'' markers.}
 \end{figure}

 \begin{figure}[H]
   \centering
   \includegraphics{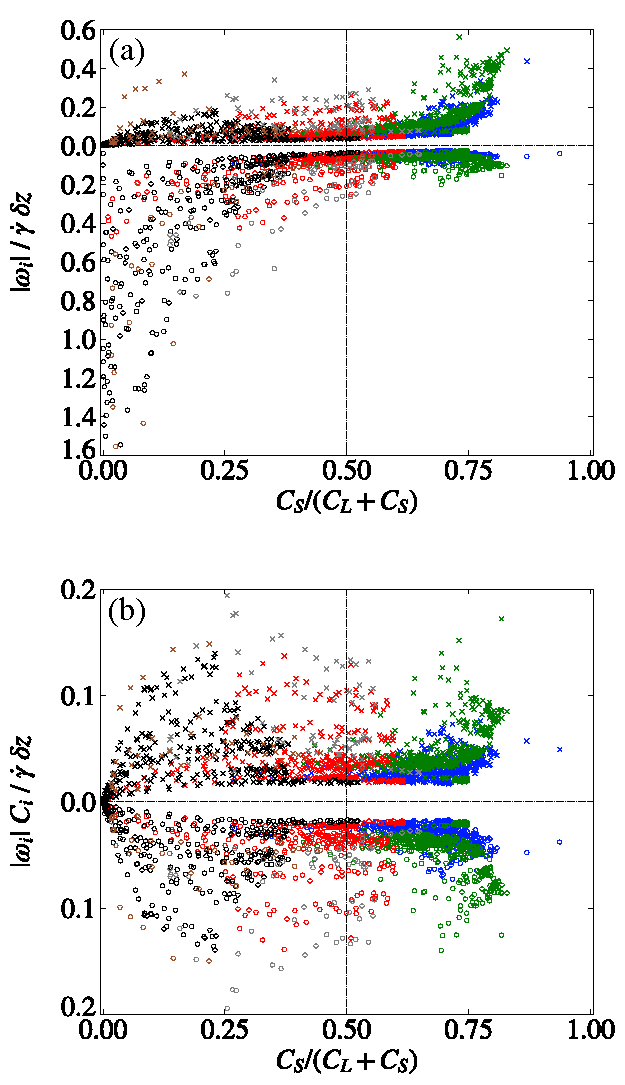 }
   \caption{\label{fig:Suplementary_S3} Magnitude of the relative species velocity $|\omega_i|$ (a) and magnitude of the flux $|\omega_i| C_i$ (b), scaled by the local shear rate $\dot \gamma\ \delta z$, as a function of the local small particles concentration $C_S/(C_L+C_S)$ in numerical simulations.
   Data correspond only to the shear zone $z<12.5d$ before the steady state. Colors represent different simulations, large particles are represented with ``x'' markers, and small particles with ``o'' markers. 
   }
 \end{figure}

 \begin{figure}[H]
   \centering
   \includegraphics{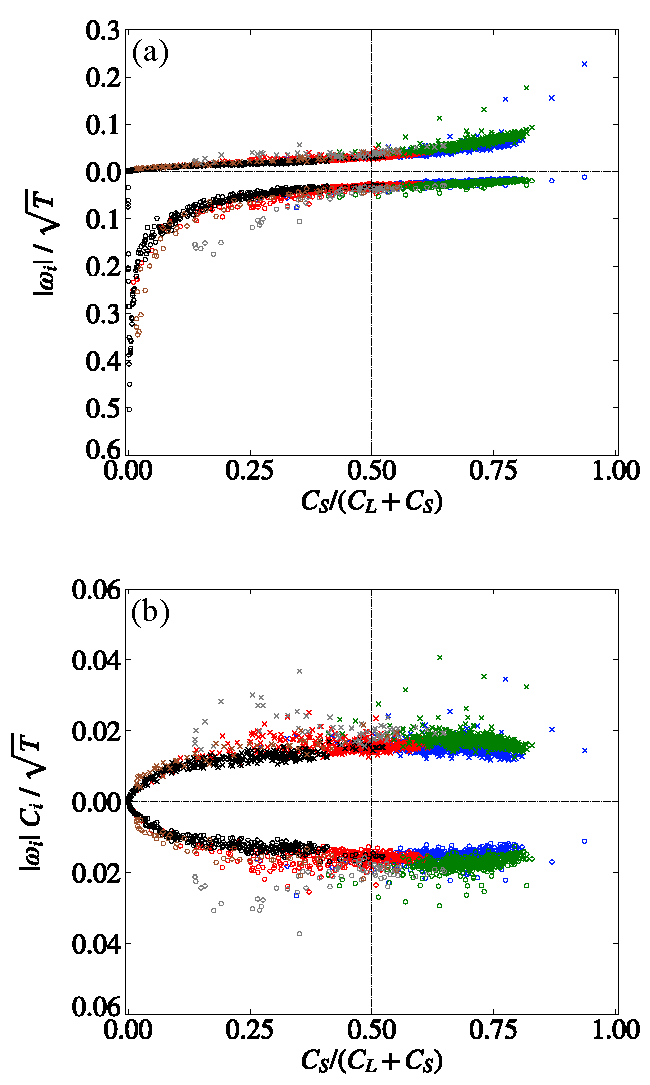 }
   \caption{\label{fig:Suplementary_S5} Magnitude of the relative species velocity $|\omega_i|$ (a) and magnitude of the flux $|\omega_i| C_i$ (b), scaled by the square root of the granular temperature $\sqrt{T}$, as a function of the local small particles concentration $C_S/(C_L+C_S)$ in numerical simulations.
   Data correspond only to the shear zone $z<12.5d$ before the steady state. Colors represent different simulations, large particles are represented with ``x'' markers, and small particles with ``o'' markers. 
   }
 \end{figure}
